\pdfoutput=1
\documentclass[12pt,a4paper]{article}

\newcommand{\munu}{\mu^-\overline{\nu}}
\newcommand{\flb}{f_{\Lambda_b}}
\newcommand{\rbs}{f_s/(f_u+f_d)}
\newcommand{\Dssz}{D_{s0}^*(2317)}
\newcommand{\Dssone}{D_{s1}(2460)}
\newcommand{\Dssonep}{D_{s1}(2536)}
\newcommand{\Lcstar}{{\it \Lambda _c}(2595)}
\newcommand{\Lcstarp}{{\it \Lambda _c}(2625)}
\usepackage{graphicx}  

\usepackage{xspace}
\usepackage{color}
\usepackage{colortbl}

\usepackage{amsmath}

\usepackage{ifthen} 

\newboolean{pdflatex}
\setboolean{pdflatex}{true} 
\newboolean{uprightparticles}
\setboolean{uprightparticles}{false} 
\usepackage{amssymb}
\usepackage{amsfonts}
\usepackage{upgreek}

\usepackage{hyperref}
\usepackage[all]{hypcap} 

\def\ux85 {UX85\xspace}

\ifthenelse{\boolean{uprightparticles}}%
{

 \def\Pnu         {\ensuremath{\upnu}\xspace}                 
                  
 \def\Ppi         {\ensuremath{\uppi}\xspace}

 \def\PDelta      {\ensuremath{\Delta}\xspace}                 
 \def\PXi      {\ensuremath{\Xi}\xspace}                 
 \def\PLambda      {\ensuremath{\Lambda}\xspace}                 
 \def\PSigma      {\ensuremath{\Sigma}\xspace}                 
 \def\POmega      {\ensuremath{\Omega}\xspace}                 
 \def\PUpsilon      {\ensuremath{\Upsilon}\xspace}                 
 
 \def\PB      {\ensuremath{\mathrm{B}}\xspace}                 
                  
 \def\PD      {\ensuremath{\mathrm{D}}\xspace}

 \def\PK      {\ensuremath{\mathrm{K}}\xspace}

 \def\Pb      {\ensuremath{\mathrm{b}}\xspace}                 
 \def\Pc      {\ensuremath{\mathrm{c}}\xspace}

 \def\Pi      {\ensuremath{\mathrm{i}}\xspace}

 \def\Ps      {\ensuremath{\mathrm{s}}\xspace}

}
{

 \def\Pnu         {\ensuremath{\nu}\xspace}                 
                  
 \def\Ppi         {\ensuremath{\pi}\xspace}

 \mathchardef\PDelta="7101
 \mathchardef\PXi="7104
 \mathchardef\PLambda="7103
 \mathchardef\PSigma="7106
 \mathchardef\POmega="710A
 \mathchardef\PUpsilon="7107
                  
 \def\PB      {\ensuremath{B}\xspace}                 
                  
 \def\PD      {\ensuremath{D}\xspace}

 \def\PK      {\ensuremath{K}\xspace}

 \def\Pb      {\ensuremath{b}\xspace}                 
 \def\Pc      {\ensuremath{c}\xspace}

 \def\Pi      {\ensuremath{i}\xspace}

 \def\Ps      {\ensuremath{s}\xspace}

}

\def\neub       {\ensuremath{\overline{\Pnu}}\xspace}

\def\neumb      {\ensuremath{\neub_\mu}\xspace}

\def\squark    {\ensuremath{\Ps}\xspace}

\def\cquark    {\ensuremath{\Pc}\xspace}

\def\bquark    {\ensuremath{\Pb}\xspace}

\def\pion  {\ensuremath{\Ppi}\xspace}

\def\pip   {\ensuremath{\pion^+}\xspace}
\def\pim   {\ensuremath{\pion^-}\xspace}

\def\kaon  {\ensuremath{\PK}\xspace}
  \def\Kbar  {\kern 0.2em\overline{\kern -0.2em \PK}{}\xspace}

\def\Kz    {\ensuremath{\kaon^0}\xspace}
\def\Kzb   {\ensuremath{\Kbar^0}\xspace}
\def\KzKzb {\ensuremath{\Kz \kern -0.16em \Kzb}\xspace}
\def\Kp    {\ensuremath{\kaon^+}\xspace}
\def\Km    {\ensuremath{\kaon^-}\xspace}

\def\KpKm  {\ensuremath{\Kp \kern -0.16em \Km}\xspace}

  \def\Dbar    {\kern 0.2em\overline{\kern -0.2em \PD}{}\xspace}
\def\D       {\ensuremath{\PD}\xspace}

\def\Dz      {\ensuremath{\D^0}\xspace}
\def\Dzb     {\ensuremath{\Dbar^0}\xspace}
\def\DzDzb   {\ensuremath{\Dz {\kern -0.16em \Dzb}}\xspace}
\def\Dp      {\ensuremath{\D^+}\xspace}
\def\Dm      {\ensuremath{\D^-}\xspace}

\def\DpDm    {\ensuremath{\Dp {\kern -0.16em \Dm}}\xspace}

\def\Ds      {\ensuremath{\D^+_\squark}\xspace}
\def\Dsp     {\ensuremath{\D^+_\squark}\xspace}
\def\Dsm     {\ensuremath{\D^-_\squark}\xspace}

\def\Dss     {\ensuremath{\D^{*+}_\squark}\xspace}

\def\B       {\ensuremath{\PB}\xspace}
  \def\Bbar    {\kern 0.18em\overline{\kern -0.18em \PB}{}\xspace}

\def\Bzb     {\ensuremath{\Bbar^0}\xspace}

\def\Bub     {\ensuremath{\B^-}\xspace}

\def\Bm      {\ensuremath{\Bub}\xspace}

\def\Bsb     {\ensuremath{\Bbar^0_\squark}\xspace}

  \def\Y#1S{\ensuremath{\PUpsilon{(#1S)}}\xspace}

\def\L {\ensuremath{\PLambda}\xspace}

\def\Lb      {\ensuremath{\L_\bquark}\xspace}

\def\Lc      {\ensuremath{\L_\cquark}\xspace}

\def\to                 {\ensuremath{\rightarrow}\xspace}

\def\qsq       {\ensuremath{q^2}\xspace}

\def\AT#1     {\ensuremath{A_{\mathrm{T}}^{#1}}\xspace}           

\def\C#1      {\ensuremath{\mathcal{C}_{#1}}\xspace}                       
\def\Cp#1     {\ensuremath{\mathcal{C}_{#1}^{'}}\xspace}                    
\def\Ceff#1   {\ensuremath{\mathcal{C}_{#1}^{\mathrm{(eff)}}}\xspace}        
\def\Cpeff#1  {\ensuremath{\mathcal{C}_{#1}^{'\mathrm{(eff)}}}\xspace}       
\def\Ope#1    {\ensuremath{\mathcal{O}_{#1}}\xspace}                       
\def\Opep#1   {\ensuremath{\mathcal{O}_{#1}^{'}}\xspace}                    

\newcommand{\tev}{\ensuremath{\mathrm{\,Te\kern -0.1em V}}\xspace}
\newcommand{\gev}{\ensuremath{\mathrm{\,Ge\kern -0.1em V}}\xspace}
\newcommand{\mev}{\ensuremath{\mathrm{\,Me\kern -0.1em V}}\xspace}
\newcommand{\kev}{\ensuremath{\mathrm{\,ke\kern -0.1em V}}\xspace}
\newcommand{\ev}{\ensuremath{\mathrm{\,e\kern -0.1em V}}\xspace}
\newcommand{\gevc}{\ensuremath{{\mathrm{\,Ge\kern -0.1em V\!/}c}}\xspace}
\newcommand{\mevc}{\ensuremath{{\mathrm{\,Me\kern -0.1em V\!/}c}}\xspace}
\newcommand{\gevcc}{\ensuremath{{\mathrm{\,Ge\kern -0.1em V\!/}c^2}}\xspace}
\newcommand{\gevgevcccc}{\ensuremath{{\mathrm{\,Ge\kern -0.1em V^2\!/}c^4}}\xspace}
\newcommand{\mevcc}{\ensuremath{{\mathrm{\,Me\kern -0.1em V\!/}c^2}}\xspace}

\newcommand{\chisq}{\ensuremath{\chi^2}\xspace}

\def\gsim{{~\raise.15em\hbox{$>$}\kern-.85em
          \lower.35em\hbox{$\sim$}~}\xspace}
\def\lsim{{~\raise.15em\hbox{$<$}\kern-.85em
          \lower.35em\hbox{$\sim$}~}\xspace}

\def\tell1  {TELL1\xspace}
\def\ukl1   {UKL1\xspace}

\newcommand{\etal}{{\slshape et al.\/}\xspace}

\begin{document}

\begin{titlepage}
\pagenumbering{roman}

\vspace*{-1.5cm}
\centerline{\large EUROPEAN ORGANIZATION FOR NUCLEAR RESEARCH (CERN)}
\vspace*{1.5cm}
\hspace*{-0.5cm}
\begin{tabular*}{\linewidth}{lc@{\extracolsep{\fill}}r}
\ifthenelse{\boolean{pdflatex}}
{\vspace*{-2.7cm}\mbox{\!\!\!\includegraphics[width=.14\textwidth]{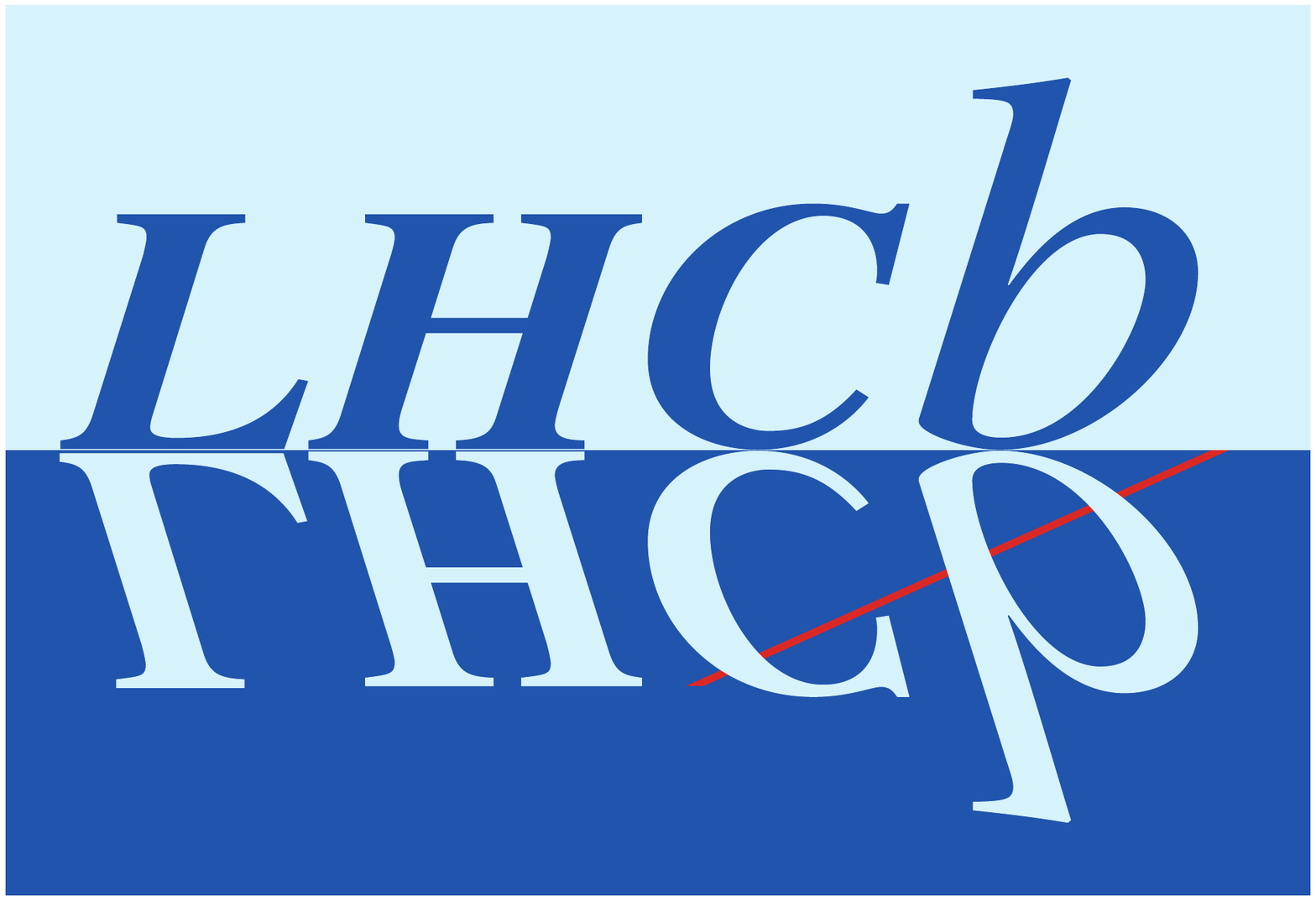}} & &}%
{\vspace*{-1.2cm}\mbox{\!\!\!\includegraphics[width=.12\textwidth]{lhcb-logo}} & & }\\
 & & CERN-PH-EP-2011-172 \\  
 & & LHCb-PAPER-2011-018 \\  
 & & \today \\ 
 & & \\
\end{tabular*}

\vspace*{4.0cm}

{\bf\boldmath\huge
\begin{center}
Measurement of $b$ hadron production fractions in 7 TeV  $pp$ collisions  
\end{center}
}


\begin{center}
The LHCb Collaboration
\footnote{Authors are listed on the following pages.}
\end{center}


\begin{abstract}
  \noindent
   Measurements of $b$ hadron production ratios in proton-proton collisions at a centre-of-mass energy of 7 TeV with an integrated luminosity of 3 pb$^{-1}$ are presented.
   We study the ratios of strange $B$ meson to light $B$ meson production $f_s/(f_u+f_d)$ and $\Lb^0$ baryon to light $B$ meson production $f_{\Lambda _b}/(f_u+f_d)$ as a function of the  charmed hadron-muon pair  transverse momentum $p_{\rm T}$ and the $b$ hadron pseudorapidity $\eta$, for $p_{\rm T}$ between 0 and 14 GeV and $\eta$ between 2 and 5. We find that $f_s/(f_u+f_d)$ is consistent with being independent of  $p_{\rm T}$ and $\eta$, and  we determine 
 $f_s/(f_u+f_d)$ = 0.134$\pm$0.004$^{+0.011}_{-0.010}$, where the first error is statistical and the second systematic. The corresponding ratio  $f_{\Lambda_b}/(f_u+f_d)$  is found to be dependent upon the transverse momentum of the charmed hadron-muon pair,   
$f_{\Lambda _b}/(f_u+f_d)=(0.404\pm 0.017 {\rm (stat)} \pm 0.027 {\rm (syst)} \pm 0.105 {\rm (Br)} )\times[1 -(0.031 \pm 0.004 {\rm (stat)}  \pm 0.003 {\rm (syst)})\times p_{\rm T} {\rm (GeV)}]$,
 where Br reflects an absolute scale uncertainty due to the poorly known branching fraction ${\cal B}(\Lc^+ \to pK^-\pi^+)$.  
  We extract the ratio of strange $B$ meson to light neutral $B$ meson production
  $f_s/f_d$  by averaging the result reported here with two previous measurements derived from the relative abundances of $\Bsb \to \Dsp \pim$ to $\Bzb \to \Dp\Km$ and $\Bzb\to \Dp\pim$. We obtain $f_s/f_d=0.267^{+0.021}_{-0.020}$. 

\end{abstract}

\vspace*{2.0cm}
\vspace{\fill}

\end{titlepage}




\cleardoublepage

\begin{flushleft}
R.~Aaij$^{23}$, 
C.~Abellan~Beteta$^{35,n}$, 
B.~Adeva$^{36}$, 
M.~Adinolfi$^{42}$, 
C.~Adrover$^{6}$, 
A.~Affolder$^{48}$, 
Z.~Ajaltouni$^{5}$, 
J.~Albrecht$^{37}$, 
F.~Alessio$^{37}$, 
M.~Alexander$^{47}$
% ...

{\it\footnotesize
$ ^{1}$Centro Brasileiro de Pesquisas F\'{i}sicas (CBPF), Rio de Janeiro, Brazil\\
% ...
$ ^{5}$Clermont Universit\'{e}, Universit\'{e} Blaise Pascal, CNRS/IN2P3, LPC, Clermont-Ferrand, France\\
$ ^{6}$CPPM, Aix-Marseille Universit\'{e}, CNRS/IN2P3, Marseille, France\\
% ...
$ ^{23}$Nikhef National Institute for Subatomic Physics, Amsterdam, Netherlands\\
% ...
$ ^{35}$Universitat de Barcelona, Barcelona, Spain\\
$ ^{36}$Universidad de Santiago de Compostela, Santiago de Compostela, Spain\\
$ ^{37}$European Organization for Nuclear Research (CERN), Geneva, Switzerland\\
% ...
$ ^{42}$H.H. Wills Physics Laboratory, University of Bristol, Bristol, United Kingdom\\
% ...
$ ^{47}$School of Physics and Astronomy, University of Glasgow, Glasgow, United Kingdom\\
$ ^{48}$Oliver Lodge Laboratory, University of Liverpool, Liverpool, United Kingdom\\
% ...
$ ^{n}$LIFAELS, La Salle, Universitat Ramon Llull, Barcelona, Spain\\
% ...
}
\end{flushleft}




\pagestyle{plain} 
\pagenumbering{arabic}


%
\section{Introduction}
The fragmentation process, in which a primary $b$ quark forms either a $b\bar{q}$ meson or a $bq_1q_2$ baryon, cannot be reliably predicted because it is driven by strong dynamics in the non-perturbative regime. Thus fragmentation functions for the various hadron species must be determined experimentally.  The LHCb experiment at the LHC  explores a unique kinematic region: it detects $b$ hadrons produced in a cone centered around the beam axis covering a region of pseudorapidity $\eta$, defined in terms of the polar angle $\theta$ with respect to the beam direction as 
$-\ln(\tan{\theta/2})$, ranging approximately between 2 and 5.   Knowledge  of the fragmentation functions allows us  to relate theoretical predictions of  the $b\bar{b}$ quark production  cross-section, derived from perturbative QCD, to the observed hadrons.  In addition, since many absolute branching fractions of $B^-$ and $\Bzb$ decays have been well measured at $e^+e^-$ colliders~\cite{PDG}, it suffices to measure the ratio of $\Bsb$ production to either $B^-$ or $\Bzb$ production to perform precise absolute $\Bsb$ branching fraction measurements.  In this paper we describe measurements of two ratios of fragmentation functions: $f_s/(f_u+f_d)$ and $\flb/(f_u+f_d)$,  where $f_q\equiv {\cal B}(b\to B_q)$ and $\flb \equiv {\cal B}(b\to \Lambda _b)$.  The inclusion of charged conjugate modes is implied throughout the paper, and we measure the average production ratios.

Previous measurements of these fractions have been made at LEP~\cite{hfag}
and at CDF~\cite{cdf:fractions}.  More recently, LHCb measured the ratio  $f_s/f_d$ using the decay modes $\Bzb\to D^+\pi^-$, $\Bzb\to D^+ K^-$, and $\Bsb \to D^+_s \pi^-$~\cite{fs:had} and theoretical input from QCD factorization~\cite{Fleischer:2010,Fleischer:2011}. Here we measure this ratio using semileptonic decays without any significant model dependence. A commonly adopted assumption is that the fractions of these different species should be the same in high energy $b$ jets originating from $Z^0$ decays and high $p_{\rm T}$ $b$ jets originating from $p\bar{p}$ collisions at the Tevatron or $pp$ collisions at LHC, based on the notion that hadronization is a non-perturbative process occurring at the scale of $\Lambda _{\rm QCD}$. Nonetheless, the results from different experiments are discrepant in the case of  the $b$ baryon fraction~\cite{hfag}.

The measurements reported in this paper are performed using the LHCb detector~\cite{lhcb:detector}, a forward spectrometer designed to study  production and decays of hadrons containing $b$ or $c$ quarks. LHCb includes a vertex detector (VELO), providing precise locations of primary $pp$ interaction vertices, and of detached vertices of long lived hadrons. The momenta of charged particles are determined using information from the VELO together with the rest of the tracking system, composed of a large area silicon tracker located before a 4 Tm dipole magnet, and a combination of silicon strip  and straw drift chamber  detectors located after the magnet. Two Ring Imaging Cherenkov (RICH) detectors are used for charged hadron identification.  Photon detection and electron identification are implemented through an electromagnetic calorimeter followed by a hadron calorimeter. A system of alternating layers of iron and chambers  provides muon identification. The two calorimeters and the muon system provide the energy and momentum information to implement a first level (L0) hardware trigger. An additional trigger level is software based, and its algorithms are tuned to the experiment's operating condition.

In this analysis we use a data sample of  3 pb$^{-1}$  collected from 7 TeV centre-of-mass energy $pp$ collisions at the LHC during 2010. The trigger selects events where a single muon is detected without biasing the impact parameter distribution of the decay products of the $b$ hadron, nor any kinematic variable relevant to semileptonic decays.  These  features reduce the systematic uncertainty in the efficiency.
Our goal is to measure two specific production ratios: that of $\Bsb$ relative to the sum of $B^-$ and $\Bzb$, and that of $\Lb^0$, relative to the sum of $\Bm$ and $\Bzb$. The sum of the $\Bzb$, $\Bm$, $\Bsb$ and $\Lb^0$ fractions does not equal one, as there is other $b$ production, namely a very small rate for $B_c^-$ mesons, bottomonia, and other $b$ baryons that do not decay strongly into $\Lb^0$, such as the $\Xi_b$. We measure relative fractions by studying the final states $D^0\munu X$, $D^+\munu X$, $\Ds\munu X$,  $\Lc^+\munu X$, $D^0 K^+\munu X$, and $D^0 p\munu X$.  We do not attempt to separate $f_u$ and $f_d$, but we measure the sum of $\Dz$ and $\Dp$ channels and correct for cross-feeds from $\Bsb$ and $\Lb^0$ decays. We assume near equality of the semileptonic decay width of all $b$ hadrons, as discussed below. Charmed hadrons are reconstructed through the modes listed in Table~\ref{tab:charms}, together with their branching fractions. We use all $\Ds \to K^-K^+\pi^+$ decays rather than a combination of the resonant $\phi\pi^+$ and $\overline{K}^{*0}K^+$
contributions, because these $\Ds$ decays cannot be cleanly isolated due to interference effects of different amplitudes.

\begin{table}
\begin{center}
\caption{Charmed hadron decay modes and branching fractions.}\label{tab:charms}
\begin{tabular}{ccc}\hline\hline
Particle & Final state & Branching fraction (\%)\\\hline
$D^0$ & $K^-\pi^+$  & 3.89$\pm$0.05~\cite{PDG} \\
$D^+$ & $K^-\pi^+\pi^+$ & 9.14$\pm$0.20~\cite{CLEODp} \\
$D_s^+$ & $K^-K^+\pi^+$  &5.50$\pm$0.27~\cite{CLEODs} \\
$\Lambda_c^+$ & $pK^-\pi^+$  & 5.0$\pm$1.3~\cite{PDG}  \\
\hline\hline
\end{tabular}
\end{center}
\end{table}

Each of these different charmed hadron plus muon final states can be populated by a combination of initial $b$ hadron states.  $\Bzb$ mesons decay semileptonically into a mixture of $D^0$ and $D^+$ mesons, while $B^-$ mesons decay predominantly into $D^0$ mesons with a smaller admixture of $D^+$ mesons. Both include a tiny component of $\Ds K$ meson pairs. $\Bsb$ mesons decay predominantly into $D_s^+$ mesons, but can also decay into $D^0K^+$ and $D^+K_S^0$ mesons; this is expected if the 
$\Bsb$ decays into a $D_s^{**}$ state that is heavy enough to decay into a $DK$ pair. In this paper we measure this contribution using $D^0K^+X\mu^-\overline{\nu}$ events. Finally, $\Lb^0$ baryons  decay mostly into $\Lc^+$ final states. We determine other contributions using $D^0pX\mu^-\overline{\nu}$ events.   We ignore the contributions of $b\to u$ decays that comprise approximately 1\% of semileptonic $b$ hadron decays~\cite{ABS}, and constitute a roughly equal portion of each $b$ species in any case.

The corrected yields for $\Bzb$ or $B^-$ decaying into $D^0\munu X$ or $D^+\munu X$, $n_{\rm corr}$, can be expressed in terms of the measured yields, $n$, as

\begin{eqnarray}
\label{eq:dzero}
n_{\rm corr}(B\to D^0 \mu)&=&\frac{1}
{{\cal{B}}(D^0\to K^-\pi^+)\epsilon(B \to D^0)}\times \\\nonumber 
&&\left[n(D^0\mu)-n(D^0K^+\mu)
\frac{\epsilon(\Bsb§ \to D^0)}{\epsilon(\Bsb \to D^0K^+)}-n(D^0p\mu)
\frac{\epsilon(\Lb^0 \to D^0)}{\epsilon(\Lb^0 \to D^0p)}\right],
\end{eqnarray}
where we use the  shorthand $n(D\mu)\equiv n(DX\munu)$.  An analogous abbreviation $\epsilon$ is used for the total trigger and detection efficiencies.  
For example, the ratio $\epsilon(\Bsb \to D^0)/\epsilon(\Bsb \to D^0K^+)$ gives the relative efficiency to reconstruct a charged $K$ in semi-muonic $\Bsb$ decays producing a $D^0$ meson. Similarly
\begin{eqnarray}
n_{\rm corr}(B\to D^+ \mu)&=&\frac{1}{\epsilon(B \to D^+)}\left[
\frac{n(D^+\mu^-)}{{\cal{B}}(D^+\to K^-\pi^+\pi^+)}-  \right. \nonumber \\  
&&\left. \frac{n(D^0K^+\mu^-)}{{\cal{B}}(D^0\to K^-\pi^+)}
\frac{\epsilon(\Bsb \to D^+)}{\epsilon(\Bsb \to D^0K^+)}  \right. \nonumber \\
&&\left.
-\frac{n(D^0p\mu^-)}{{\cal{B}}(D^0\to K^-\pi^+)}\frac{\epsilon(\Lb \to D^+)}{\epsilon(\Lb \to D^0p)}\right].\label{eq:dp}
\end{eqnarray}
Both the $\Dz X\munu$ and the $\Dp X \munu$ final states contain small components of cross-feed from $\Bsb$ decays to $\Dz K^+ X\munu$ and  to $D^+ \Kz X\munu$. These components are accounted for by the two decays $\Bsb\to D_{s1}^+ X\munu$ and $\Bsb\to D_{s2}^{*+} X\munu$ as reported in a recent LHCb publication~\cite{Aaij:2011ju}.  The third terms in Eqs.~\ref{eq:dzero} and \ref{eq:dp} are  due to a similar small cross-feed from $\Lb^0$ decays.

The number of $\Bsb$ resulting in $\Ds X\munu$ in the final state is given by
\begin{eqnarray}
 \label{eq:bsds}
n_{\rm corr}(\Bsb\to\Ds \mu)&=& \frac{1}{\epsilon(\Bsb\to\Ds)}\left[\frac{n(\Ds \mu)}{{\cal B}(\Ds\to \Kp\Km\pip)} - \right. \\ 
& &\left. N(\Bzb+B^-){\cal B}(B\to \Ds K\mu)\epsilon(\bar{B}\to\Ds K\mu)\right] , \nonumber
\end{eqnarray}
where the last term subtracts yields of $\Ds K X\munu$ final states originating from $\Bzb$ or $B^-$ semileptonic decays, and $N(\Bzb+B^-)$ indicates the total number of $\Bzb$ and $B^-$ produced. We derive this correction using the  branching fraction ${\cal B}(B\to D_s^{(*)+}K\mu\nu)=(6.1\pm 1.2)\times 10^{-4}$~\cite{BtoDsK} measured by the BaBar experiment.  
In addition, $\Bsb$ decays semileptonically into $DKX\munu$, and thus we need to add to Eq.~\ref{eq:bsds}
\begin{equation}
n_{\rm corr}(\Bsb\to DK\mu)= 2 \frac{n(\Dz K^+\mu)}{{\cal B}(\Dz\to K^-\pip)\epsilon(\Bsb\to D^0K^+\mu)} , \label{eq:bsdk}
\end{equation}
where, using  isospin symmetry, the factor of 2 accounts for $\Bsb\to D\Kzb X\munu$ semileptonic decays.

The equation for the ratio $\rbs$ is 
\begin{equation}
\label{eq:frac-final}
\frac{f_s}{f_u+f_d}=\frac{n_{\rm corr}(\Bsb\to D\mu)}{n_{\rm corr}(B\to D^0 \mu)+n_{\rm corr}(B\to D^+ \mu)}
\frac{\tau_{B^-}+\tau_{\Bzb}}{2\tau_{\Bsb}} .
\end{equation}
where $\Bsb\to D\mu$ represents  $\Bsb$ semileptonic decays to a final charmed hadron, given by the sum of the contributions shown in Eqs. \ref{eq:bsds} and \ref{eq:bsdk}, and the symbols $\tau _{B_i}$ indicate the $B_i$ hadron lifetimes, that are all well measured~\cite{PDG}.  We use the average $\Bsb$ lifetime, 1.472$\pm$0.025 ps~\cite{PDG}. This equation assumes equality of the semileptonic widths of all the $b$ meson species. This is a reliable assumption, as corrections in HQET arise only to order 1/$m_b^2$ and the SU(3) breaking correction is quite small, of the order of 1\%~\cite{wise,uraltsev,mannel}. 

The $\Lb ^0$ corrected yield is derived in an analogous manner. We determine
\begin{equation}
n_{\rm corr}(\Lb^0\to D\mu)=\frac{n(\Lambda_c^+\mu^-)}{{\cal{B}}(\Lambda_c^+\to pK^-\pi^+)\epsilon(\Lb^0 \to \Lc^+)}
+2\frac{n(D^0p\mu^-)}{{\cal{B}}(D^0\to K^-\pi^+)\epsilon(\Lb^0 \to D^0 p)},\label{eq:lb}
\end{equation}
where $D$ represents a generic charmed hadron, and extract the
$\Lambda_b^0$ fraction using
\begin{equation}
\label{eq:frac-final-lb}
\frac{f_{\Lambda_b}}{f_u+f_d}= \frac{n_{\rm corr}(\Lambda_b^0\to D \mu)}{n_{\rm corr}(B\to D^0 \mu)+n_{\rm corr}(B\to D^+ \mu)}
\frac{\tau_{B^-}+\tau_{\Bzb}}{2\tau_{\Lambda_b^0}} (1-\xi).
\end{equation}
Again,  we assume near equality of the semileptonic widths of different $b$ hadrons, but we apply a small adjustment  $\xi=4\pm$2\%, to account for the chromomagnetic correction, affecting $b$-flavoured mesons but not $b$ baryons~\cite{wise,uraltsev,mannel}. The uncertainty is evaluated with very conservative assumptions for all the parameters of the heavy quark expansion.

\section{Analysis method}

To isolate a sample of $b$ flavoured hadrons with low backgrounds, we match charmed hadron candidates with tracks identified as muons. Right-sign (RS) combinations have the sign of the charge of the muon being the same as the charge of the kaon in $\Dz$, $D^+$, or $\Lc^+$ decays, or the opposite charge of the pion in $\Ds$ decays, while wrong-sign (WS) combinations comprise combinations with opposite charge correlations. WS events are useful to estimate certain backgrounds.
This analysis follows our previous investigation of $b\to D^0 X \mu^-\overline{\nu}$~\cite{1stpaper}.  We consider events where a well-identified muon with momentum greater than 3 GeV and transverse momentum greater than 1.2 GeV is found. Charmed hadron candidates are formed from hadrons with momenta greater than 2 GeV and transverse momenta greater than 0.3 GeV, and we require that the average transverse momentum of the hadrons forming the candidate be greater than 0.7 GeV. Kaons, pions, and protons are identified using the RICH system. The impact parameter (IP), defined as the minimum distance of approach of the track with respect to the primary vertex, is used to select tracks coming from charm decays. We require that the $\chisq$, formed by using the hypothesis that each track's  IP is equal to 0, is greater than 9. Moreover, the selected tracks must be consistent with coming from a common vertex:  the $\chisq$ per number of degrees of freedom of the vertex fit must be smaller than 6. In order to ensure that the charm vertex is distinct from the primary $pp$ interaction  vertex,  we require that the  $\chisq $, based on the hypothesis that the decay flight distance from the primary vertex is zero, is greater than 100.

Charmed hadrons and muons are combined to form a partially reconstructed $b$ hadron by requiring that they come from a common vertex, and that the cosine of the angle between the momentum of the charmed hadron and muon pair and the line from the $D\mu$ vertex to the primary vertex be greater than 0.999. As the charmed hadron is a decay product of the $b$ hadron, we require that the difference in $z$ component of the decay vertex of the charmed hadron candidate and that of the beauty candidate  be greater than 0. We explicitly require that the $\eta$ of the $b$ hadron candidate be between 2 and 5. We measure $\eta$ using the line
defined by connecting the primary event vertex and the vertex formed by the $D$ and the
$\mu $. Finally, the invariant mass of the charmed hadron and muon system must be between 3 and 5 GeV for $D^0\mu^-$ and $D^+\mu^-$ candidates, between 3.1 and 5.1 GeV for $\Ds\mu^-$ candidates, and between 3.3 and 5.3 GeV for $\Lc^+\mu^-$ candidates.

We perform our analysis in a grid of 3 $\eta$ and 5 $p_{\rm T}$ bins, covering the range $2< \eta < 5$ and $p_{\rm T}\le 14$ GeV.
The $b$ hadron signal is separated from various sources of background by studying the two dimensional distribution of charmed hadron candidate invariant mass and ln(IP/mm).  This approach allows us to determine the background coming from false combinations under the charmed hadron signal mass peak directly. The study of the ln(IP/mm) distribution allows the separation of prompt charm decay candidates from charmed hadron daughters of $b$ hadrons~\cite{1stpaper}.  We refer to these samples as Prompt and Dfb respectively.

\subsection{Signal extraction}

We describe the method used to extract the charmed hadron-$\mu$ signal by using the $\Dz X\munu$ final state as an example; the same procedure is applied to the final states $\Dp X\munu$, $\Ds X\munu$,    and $\Lc^+ X\munu$. We perform unbinned
extended maximum likelihood fits to the two-dimensional distributions in $K^-\pi^+$ invariant
mass over a region extending $\pm$80 MeV from the $D^0$ mass peak, and ln(IP/mm).  The parameters of the IP distribution of the Prompt sample are found by examining directly produced charm~\cite{1stpaper} whereas a shape derived from  simulation is used for the Dfb component. 
\begin{figure}[htb]
\begin{center}
\includegraphics[width=5.9 in]{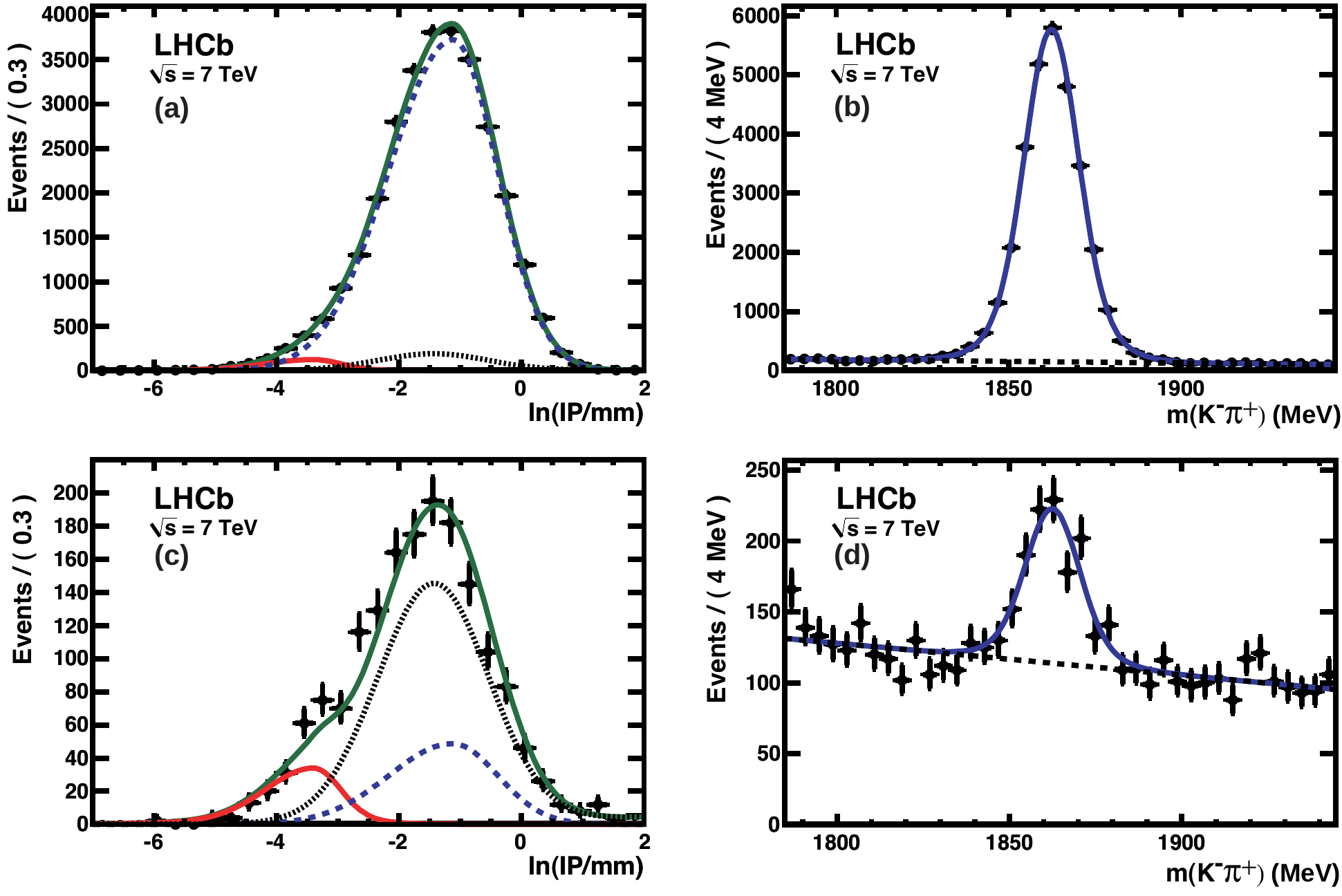}
\end{center}
\caption{The logarithm of the IP distributions for (a) RS and (c) WS $D^0$ candidate combinations with a
muon. The dotted curves show the false $D^0$ background, the small red-solid curves the Prompt yields, the dashed curves the Dfb
signal, and the larger green-solid curves the total yields. The invariant $K^-\pi^+$ mass spectra for (b) RS combinations and (d) WS combinations are also shown.
 } \label{D0-all}
\end{figure}

\begin{figure}[hbt]
\begin{center}
\includegraphics[width=5.9 in]{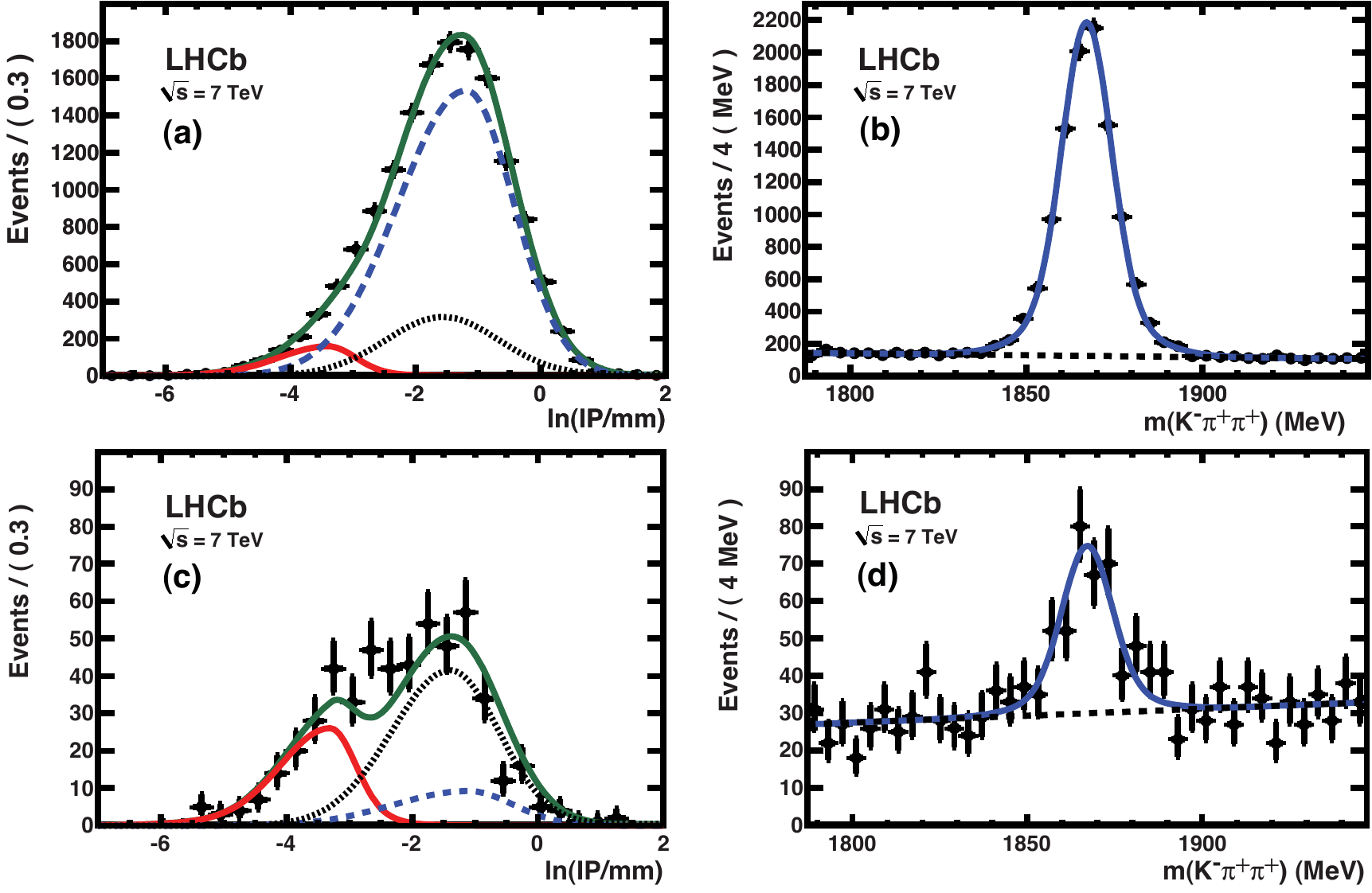}
\end{center}
\caption{The logarithm of the IP distributions for (a) RS and (c) WS $D^+$ candidate combinations with a
muon. The grey-dotted curves show the false $D^+$  background, the small red-solid curves the Prompt yields, the blue-dashed curves the Dfb
signal, and the larger green-solid curves the total yields. The invariant $K^-\pi^+\pi^+$ mass spectra for (b) RS combinations  and (d) WS combinations  are also shown.
 } \label{Dp-overall}
\end{figure}

An example fit for $D^0\munu X$, using the whole  $p_{\rm T}$ and $\eta$ range, is shown in Fig.~\ref{D0-all}. The fitted yields for RS are  27666$\pm$187 Dfb,
695$\pm$43
Prompt, and  1492$\pm$30
false $D^0$  combinations, inferred from the fitted yields in the sideband  mass regions, spanning the intervals between 35 and 75 MeV from the signal peak on both sides.  For WS we find 362$\pm$39
Dfb, 187$\pm$18 Prompt, 
and 1134$\pm$19
false $D^0$ combinations. The RS yield includes  a background of around 0.5\% from incorrectly identified $\mu$ candidates.  As this paper focuses on ratios of yields, we do not subtract this component. Figure~\ref{Dp-overall} shows the corresponding fits for the $\Dp X\munu$ final state. The fitted yields consist of   9257$\pm$110 Dfb events,
362$\pm$34
Prompt, and  1150$\pm$22
false $D^+$ combinations. For WS we find 77$\pm$22
Dfb, 139$\pm$14 Prompt
and 307$\pm$10
false $D^+$ combinations.

\begin{figure}[hbt]
\begin{center}
\includegraphics[width=5.9 in]{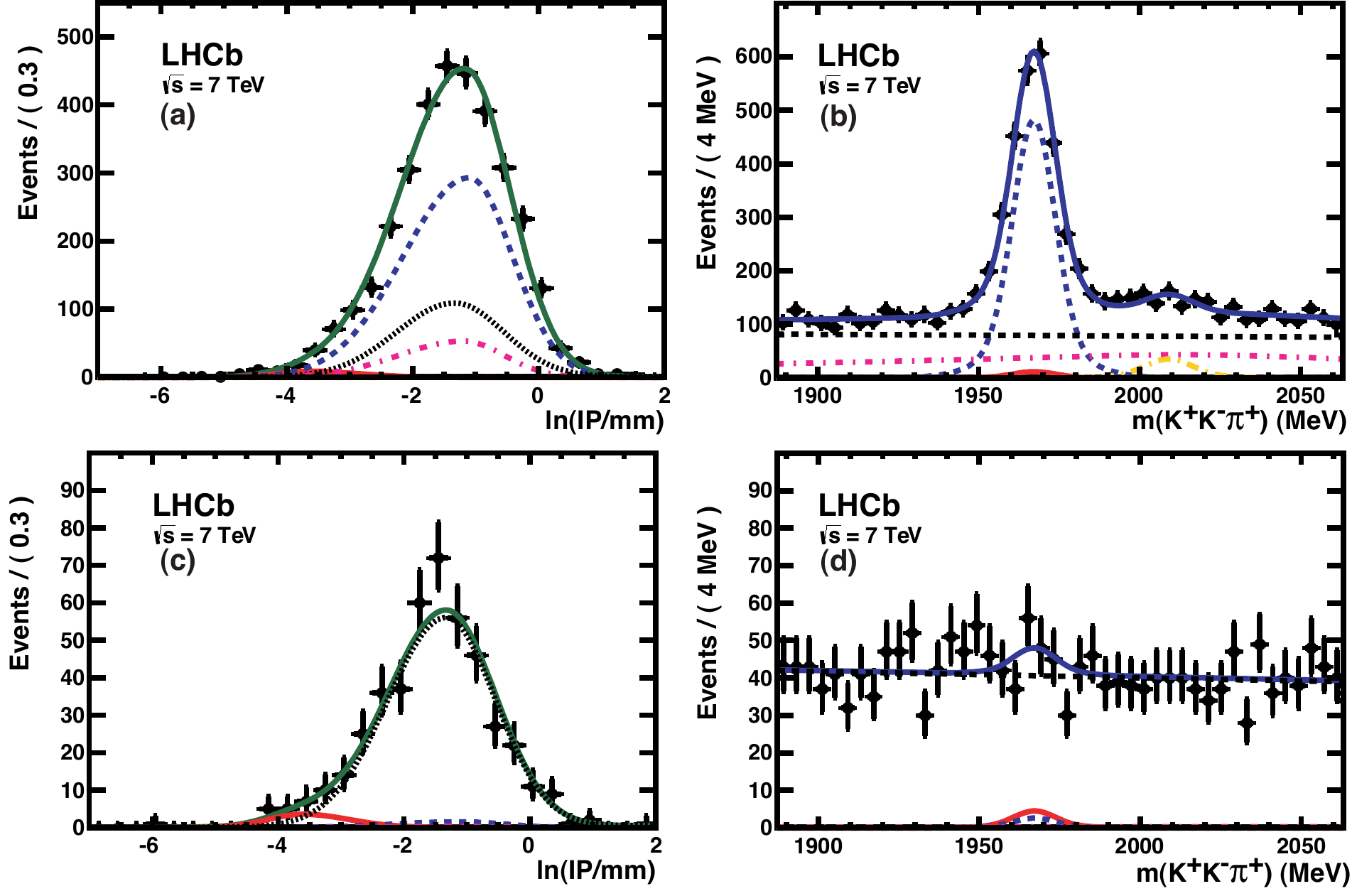}
\end{center}
\caption{The logarithm of the IP distributions for (a) RS and (c) WS $\Ds$ candidate combinations with a
muon. The grey-dotted curves show the false $\Ds$ background, the small red-solid curves the Prompt yields, the blue-dashed curves the Dfb
signal, the purple dash-dotted curves represent the background originating from $\Lc^+$ reflection, and the larger green-solid curves the total yields. The invariant $K^-K^+\pi^+$ mass spectra for RS combinations (b) and WS combinations (d) are also shown.
 } \label{Ds-overall}
\end{figure}

The analysis for the $D_s^+ X\munu$ mode follows in the same manner. Here, however, we are concerned about the reflection from $\Lambda_c^+\to pK^-\pi^+$ where the proton is taken to be a kaon, since we do not impose an explicit proton veto. Using such a veto would lose 30\% of the signal and also introduce a systematic error. We choose to model separately this particular background. 
We add a probability density function (PDF) determined from simulation to model this, and the level is allowed to float within the estimated error on the size of the background.  The small peak near 2010 MeV in Fig.~\ref{Ds-overall}(b) is due to $D^{*+}\to\pi^+D^0,D^0\to K^+K^-$. We explicitly include this term in the fit, assuming the shape to be the same as for the $\Ds$ signal, and we obtain 4$\pm$1 events in the RS signal region and no events in the WS signal region. The measured yields in the RS sample  are 2192$\pm$64 Dfb, 63$\pm$16 Prompt, 985$\pm$145 false $\Ds$ background, and 387$\pm$132 $\Lambda _c^+$ reflection background. The corresponding yields in the WS sample are 13$\pm$19, 20$\pm$7, 499$\pm$16,  and 3$\pm$3 respectively.  Figure~\ref{Ds-overall} shows the fit results.

\begin{figure}[bt]
\begin{center}
\includegraphics[width=5.9 in]{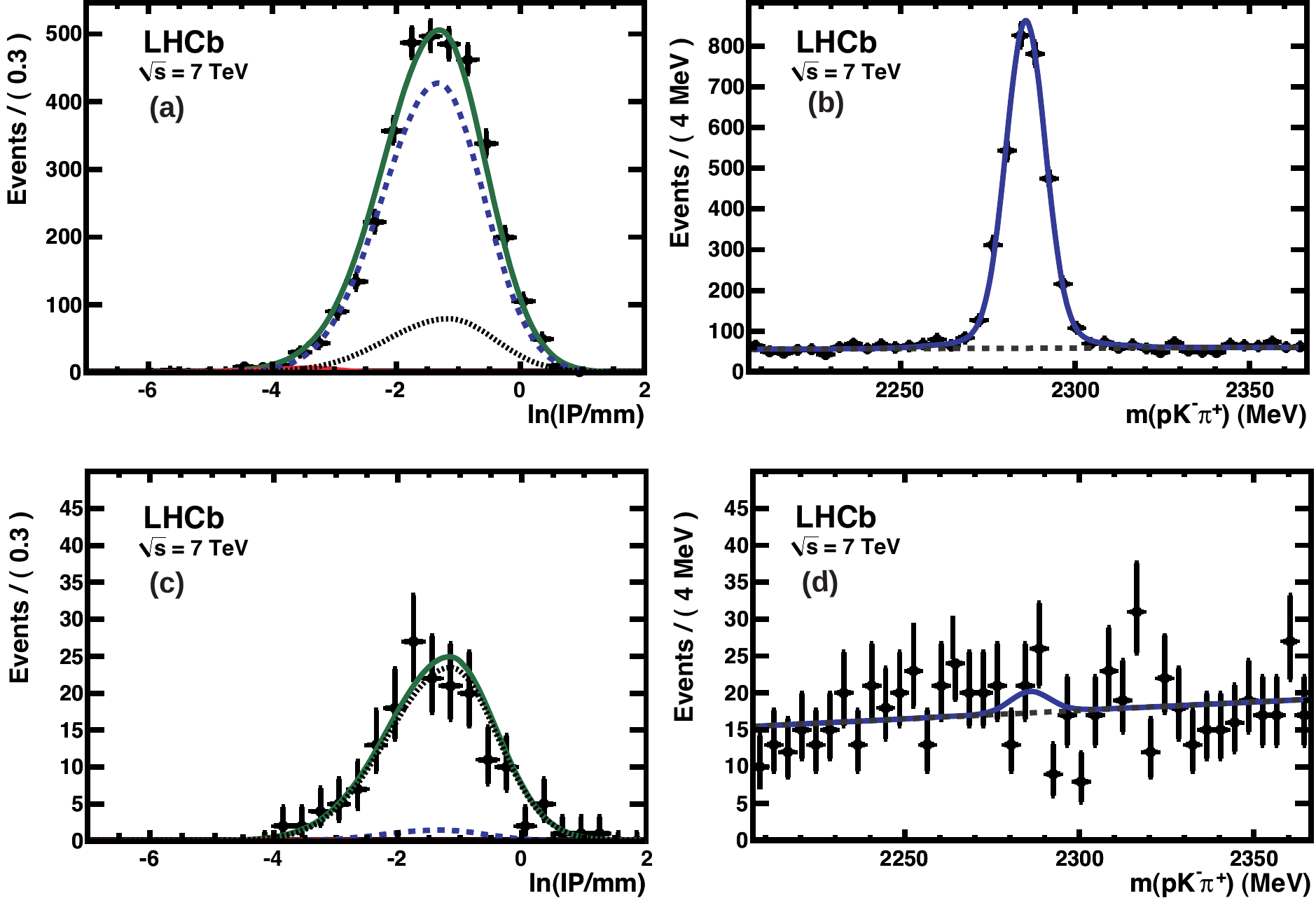}
\end{center}
\caption{The logarithm of the IP distributions for (a) RS and (c) WS $\Lambda_c^+$ candidate combinations with a
muon. The grey-dotted curves show the false $\Lambda_c^+$ background, the small red-solid curves the Prompt yields, the blue-dashed curves the Dfb
signal, and the larger green-solid curves the total yields. The invariant $p K^-\pi^+$ mass spectra for RS combinations (b) and WS combinations (d) are also shown.
 } \label{Lc-overall}
\end{figure}

The last final state  considered is $\Lc^+ X\munu $. Figure~\ref{Lc-overall} shows the data and fit components to the ln(IP/mm) and $pK^-\pi^+$ invariant mass combinations for events with  $2<\eta<5$. This fit gives 3028$\pm$112 RS Dfb events,   43$\pm$17 RS Prompt events, 589$\pm$27 RS false $\Lc ^+$ combinations, 
  9$\pm$16 WS Dfb events, 
    0.5$\pm$4 WS Prompt events, and
 177$\pm$10 WS false $\Lc^+$ combinations.
\begin{figure}[hbt]
\begin{center}
\includegraphics[width=5.9 in]{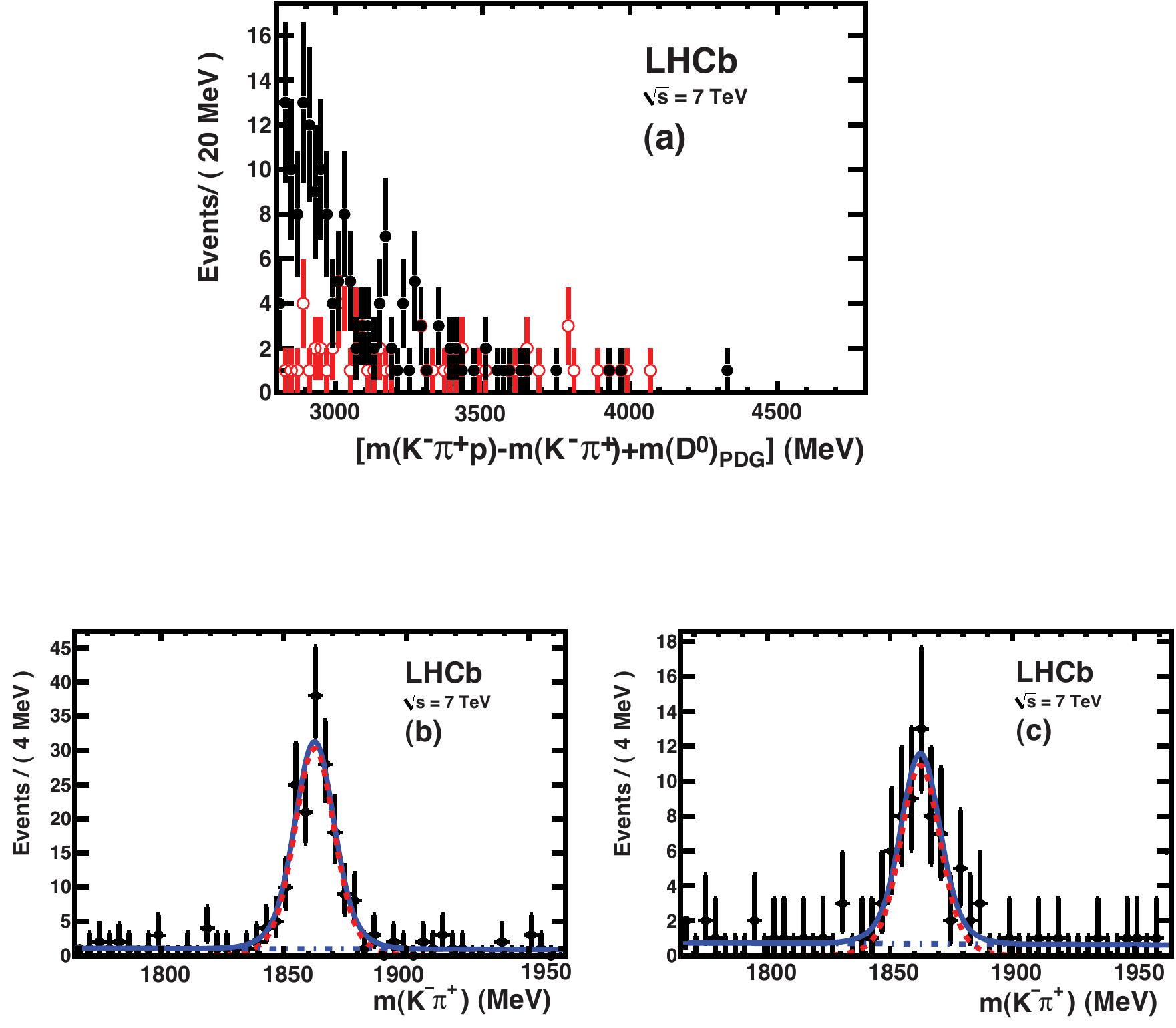}
\end{center}
\caption{(a) Invariant mass of $D^0 p$ candidates that vertex with each other and together with a RS muon (black closed points) and for a $\overline{p}$ (red open points) instead of a $p$; (b) fit to $D^0$ invariant mass for RS events with the invariant mass of $D^0 p$ candidate in the signal mass difference window;  (c) fit to $D^0$ invariant mass for WS events with the invariant mass of $D^0 p$ candidate in the signal mass difference window.}
 \label{m_d0p}
\end{figure}


The $\Lb^0$ may  also decay into $\Dz pX \munu$. We search for these decays by requiring the presence of a track well identified as a proton and  detached from any primary vertex. The resulting $D^0 p$ invariant mass distribution is shown in Fig.~\ref{m_d0p}. We also show the combinations that cannot arise from $\Lambda_b^0$ decay, namely those with $D^0 \overline{p}$ combinations. There is a clear excess of RS over WS combinations especially near threshold. Fits to the $K^-\pi^+$ invariant mass in the $[m(K^-\pi^+p)-  m(K^-\pi^+) +m(D^0)_{\rm PDG}]$ region shown in Fig.~\ref{m_d0p}(a) give 154$\pm$13 RS events and 55$\pm$8 WS events.  In this case, we use the WS yield for background subtraction, scaled by the RS/WS background ratio determined with a MC simulation including $(B^- +\Bzb\to \Dz X \munu)$ and generic $b\overline{b}$ events. This ratio is found to be 1.4$\pm$0.2. Thus, the net signal is 76$\pm$17$\pm$11, where the last error reflects the uncertainty in the 
ratio between RS and WS background. 

\clearpage

\subsection{Background studies}

Apart from false $D$ combinations, separated from the signal by the two-dimensional fit described above, there are also physical background sources that affect
the RS Dfb samples, and originate from $b\overline{b}$ events, which are studied with a MC simulation.  In the meson case, the background mainly comes from $b\to DDX$ with one of the $D$ mesons decaying semi-muonically, and from combinations of tracks from the $pp\to b\bar{b}X$ events, where one $b$ hadron decays into a $D$ meson and the other $b$ hadron decays semi-muonically.
The background fractions are (1.9$\pm$0.3)\% for $D^0 X \munu$, (2.5$\pm$0.6)\% for $D^+ X \munu$, and (5.1$\pm$1.7)\% for $D_s^+ X \munu$.  
The main background component for $\Lb^0$ semileptonic decays  is  $\Lb^0$ decaying into $\Dsm \Lc^+$, and the $\Dsm$  decaying semi-muonically.   Overall,  we find a very small background rate of (1.0$\pm$0.2)\%, where the error reflects only the statistical uncertainty in the simulation. We correct the candidate $b$ hadron yields in the signal region with the predicted background fractions. A conservative  3\% systematic uncertainty in the background subtraction is assigned to reflect  modelling uncertainties.

\begin{figure}[bt]
\begin{center}
\includegraphics[width=6.5in]{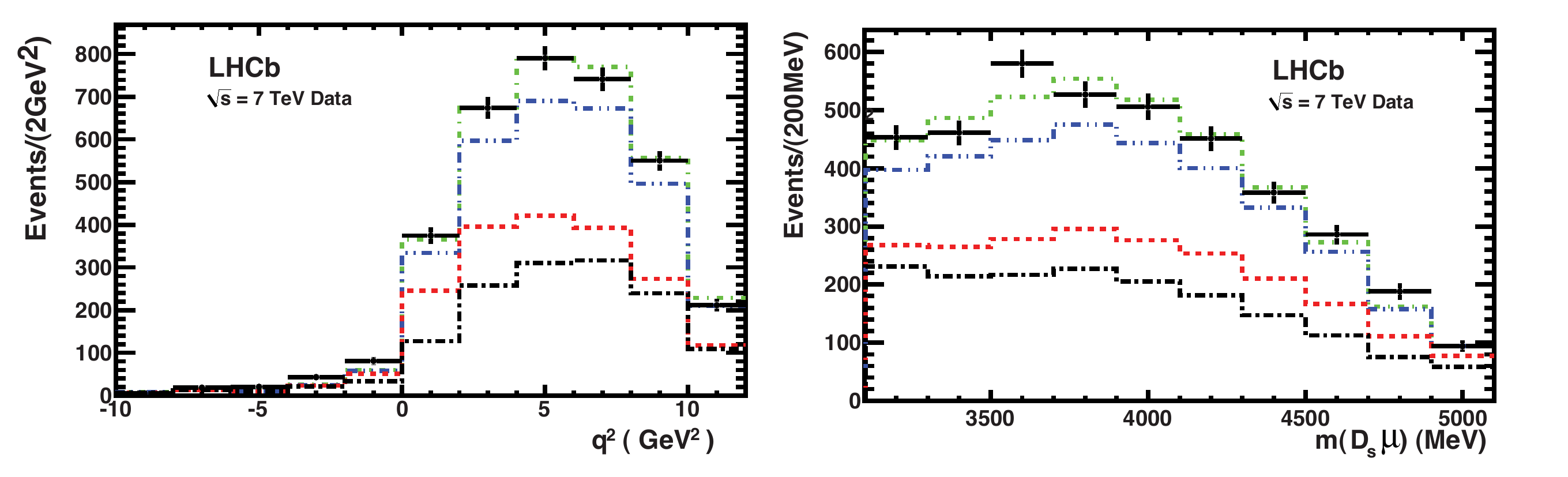}
\end{center}
\caption{Projections of the two-dimensional fit  to the $q^2$   and $m(\Ds\mu)$ distributions of semileptonic decays including a $\Ds$ meson.  The $D_s^*/D_s$ ratio has been fixed to the measured $D^*/D$ ratio in light $B$ decays  (2.42$\pm$0.10),  and the background contribution is obtained using  the sidebands in the $K^+K^-\pi^+$ mass spectrum. The different components are stacked: the background is represented by a black dot-dashed line, $\Ds$ by a red dashed line, $D_s ^{*+}$ by a blue dash-double dotted line and $\D_s^{**+}$ by a green dash-dotted line.} \label{Fit_Bs}
\end{figure}

\subsection{Monte Carlo simulation and efficiency determination}
In order to estimate the detection efficiency, we need some knowledge of the  different final states which contribute to the Cabibbo favoured semileptonic width, as some of the selection criteria affect final states with distinct masses and quantum numbers differently.
Although much is known about the $\Bzb$ and $\Bm$ semileptonic decays,  information on the corresponding $\Bsb$ and $\Lb^0$ semileptonic decays is rather sparse. In particular, the hadronic composition of the final states in $\Bsb$ decays is poorly known~\cite{Aaij:2011ju}, and only a study from CDF provides some constraints on the branching ratios of final states dominant in the corresponding $\Lb^0$ decays~\cite{cdf:lbexcl}.  

In the case of the $\Bsb \to \Ds$ semileptonic decays, we assume that the final states are $\Ds$, $\Dss$, $\Dssz^+$,  $\Dssone^+$, and $\Dssonep^+$. States above $DK$ threshold decay predominantly into $D^{(*)}K$ final states. We model the decays to the final states $\Ds \munu$ and $\Dss \munu$ with HQET form factors using normalization coefficients derived from studies of the corresponding $\Bzb$ and $\Bm$ semileptonic decays~\cite{PDG},  while we use the ISGW2 form factor model~\cite{ISGW2} to describe final states including higher mass resonances. 

In order to determine the ratio between the different hadron species in the final state,  we use the measured kinematic distributions of  the quasi-exclusive process $\Bsb\to\Ds\munu X$.  To reconstruct the squared invariant mass of the $\munu$ pair ($\qsq$), we exploit the measured direction of the $b$ hadron momentum, which, together with energy and momentum conservation, assuming no missing particles other than the neutrino, allow the reconstruction of the $\nu$ 4-vector, up to a two-fold ambiguity, due to its unknown orientation with respect to the $B$ flight path in its rest frame. We choose the solution corresponding to the lowest $b$ hadron momentum.   This method works well when there are no missing particles, or when the missing particles are soft, as in the case when the charmed system is a $D^*$ meson. We then perform a two-dimensional fit to the  $q^2$ versus $m(\mu\Ds)$ distribution.  Figure~\ref{Fit_Bs} shows  stacked histograms of the $\Ds$, $\Dss$, and $D_s^{**+}$ components. In the fit we constrain the ratio ${\cal B}(\Bsb\to \Dss\munu )/{\cal B}(\Bsb\to \Ds\munu) $ to be equal to the average $D^*\munu/D\munu$ ratio in semileptonic $\Bzb$ and $B^-$ decays (2.42$\pm$0.10)~\cite{PDG}. This constraint reduces the uncertainty of one $D^{**}$ fraction. We have also performed fits removing this assumption, and the variation between the different components is used to assess the modelling systematic uncertainty.

\begin{figure}[hbt]
\centering
\includegraphics[width=6in]{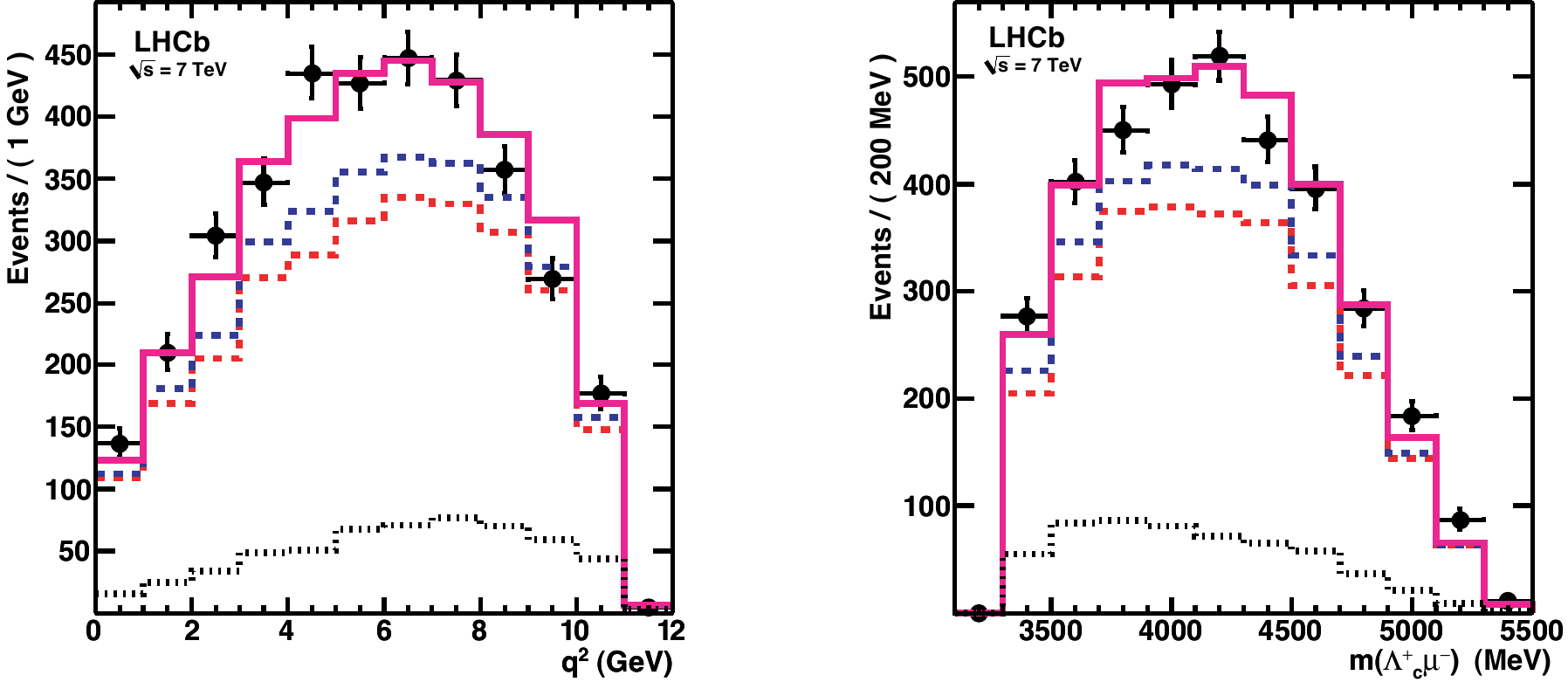}
\caption{Projections of the two-dimensional fit  to the $q^2$   and $m(\Lc^+\mu^-)$ distributions of semileptonic decays including a $\Lc^+$ baryon. The different components are stacked:
the dotted line represents the combinatoric background, the bigger dashed line (red) represents the $\Lc^+\munu$ component, the smaller dashed line (blue) the $\Lcstar^+$, and the solid line represents the $\Lcstarp^+$ component.  The $\Lcstar^+/\Lcstarp^+$ ratio is fixed to its predicted value, as described in the text.} \label{fig:Lambdab:ff}
\end{figure}

A similar procedure is applied to the $\Lc^+\mu^-$ sample and the results are shown in Fig.~\ref{fig:Lambdab:ff}. In this case we consider three final states, $\Lc^+\munu$, $\Lcstar^+\munu$, and $\Lcstarp^+\munu$, with form factors from the model of Ref.~\cite{lb-ff}. We constrain the two highest mass hadrons to be produced in the ratio predicted by this  theory. 

\begin{figure}[hbt]
\begin{center}
\includegraphics[width=3 in]{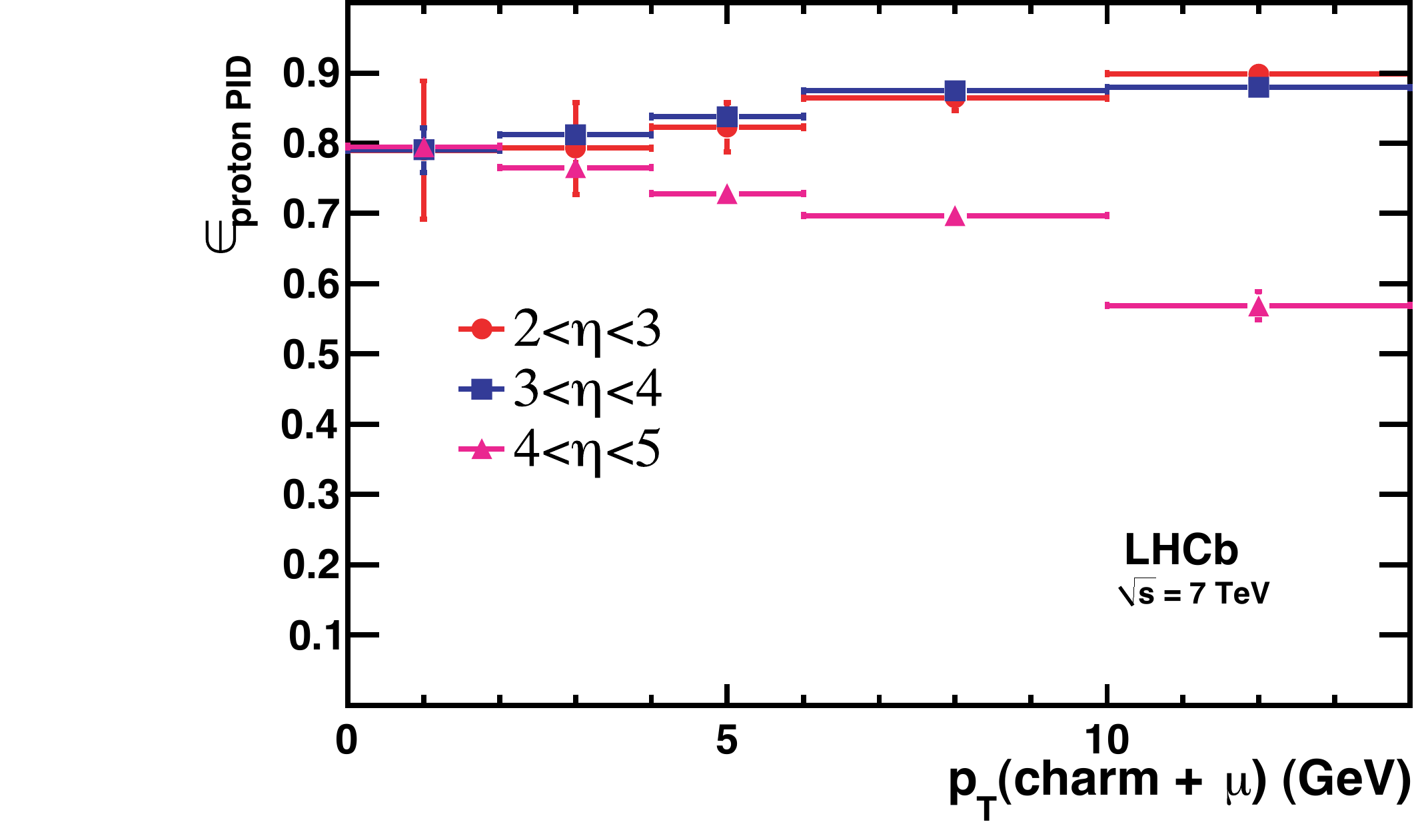}
\caption{Measured proton identification efficiency as a function of the  $\Lc^+\mu^-$ $p_{\rm T}$  for $2<\eta < 3$, $3<\eta <4$, $4<\eta < 5$ respectively, and for the selection criteria used in the $\Lc^+\to pK^-\pi^+$ reconstruction.}
\label{fig:peff}
\end{center}
\end{figure}

The measured pion, kaon and proton identification efficiencies are determined using $K_{\rm S}^0$, $D^{*+}$, and $\Lambda^0$ calibration samples where $p$, $K$, and $\pi$ are selected without utilizing the particle identification criteria. The efficiency is obtained by fitting simultaneously the invariant mass distributions of events either passing or failing the identification requirements. Values are obtained in bins of the particle $\eta$ and $p_{\rm T}$, and these efficiency matrices are applied to the MC simulation. Alternatively, the particle identification efficiency can be determined by using the measured efficiencies and combining them with weights proportional to the fraction of particle types with a given $\eta$ and $p_{\rm T}$ for each $\mu$ charmed hadron pair  $\eta$ and $p_{\rm T}$ bin. The overall efficiencies obtained with these two methods are consistent.  An example of the resulting particle identification efficiency as a function of the $\eta$ and $p_{\rm T}$  of  the  $\Lc^+\mu^-$ pair is shown in Fig.~\ref{fig:peff}. 

\begin{figure}[htb]
\begin{center}
\includegraphics[width=3in]{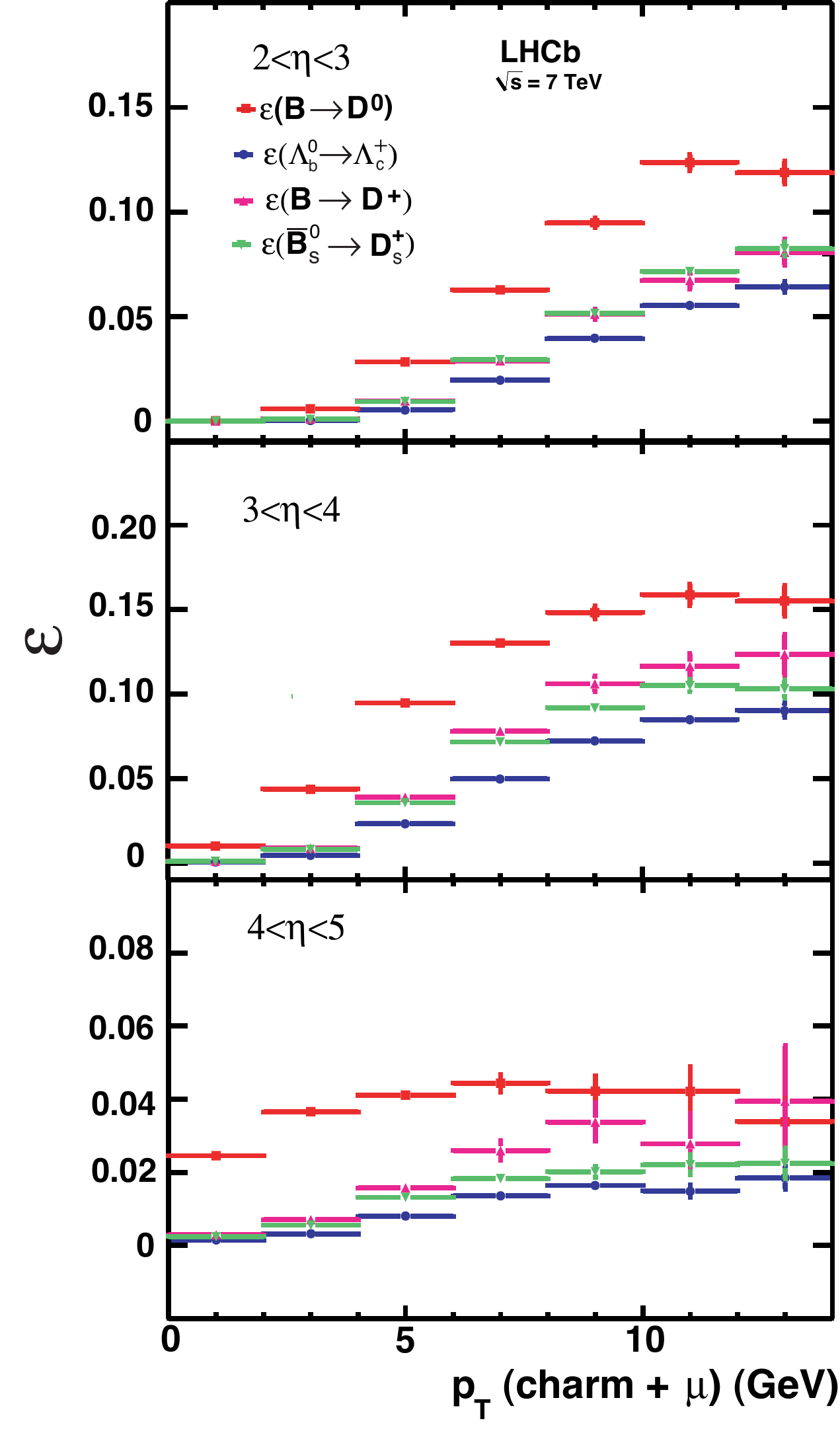}
\end{center}
\caption{Efficiencies for $\Dz\munu X$, $\Dp\munu X$, $\Ds\munu X$, $\Lc^+\munu X$ as a function of  $\eta$ and $p_{\rm T}$.} \label{eff-pt-eta}
\end{figure}

As the functional forms of the fragmentation ratios in terms of $p_{\rm T}$ and $\eta$ are not known,   we determine the efficiencies for the final states studied as a function of $p_{\rm T}$ and $\eta$ within the LHCb acceptance. Figure~\ref{eff-pt-eta} shows the results. 
\section{\boldmath Evaluation of the ratios ${f_s/(f_u+f_d)}$ and  ${f_{\Lambda _b}/(f_u+f_d)}$ }
Perturbative QCD calculations lead us to expect the ratios $f_s/(f_u+f_d)$ and $f_{\Lambda _b}/(f_u+f_d)$ to be independent of $\eta$, while a possible dependence upon the $b$ hadron transverse momentum $p_{\rm T}$ is not ruled out, especially for ratios involving baryon species~\cite{michelangelo}. Thus we determine these fractions in different $p_{\rm T}$ and $\eta$ bins.  For simplicity, we use the transverse momentum of the charmed hadron-$\mu$ pair as the $p_{\rm T}$ variable, and do not try to unfold the $b$ hadron transverse momentum. 

\begin{figure}[hbt]
\begin{center}$
\begin{array}{cc}
\includegraphics[width=6 in]{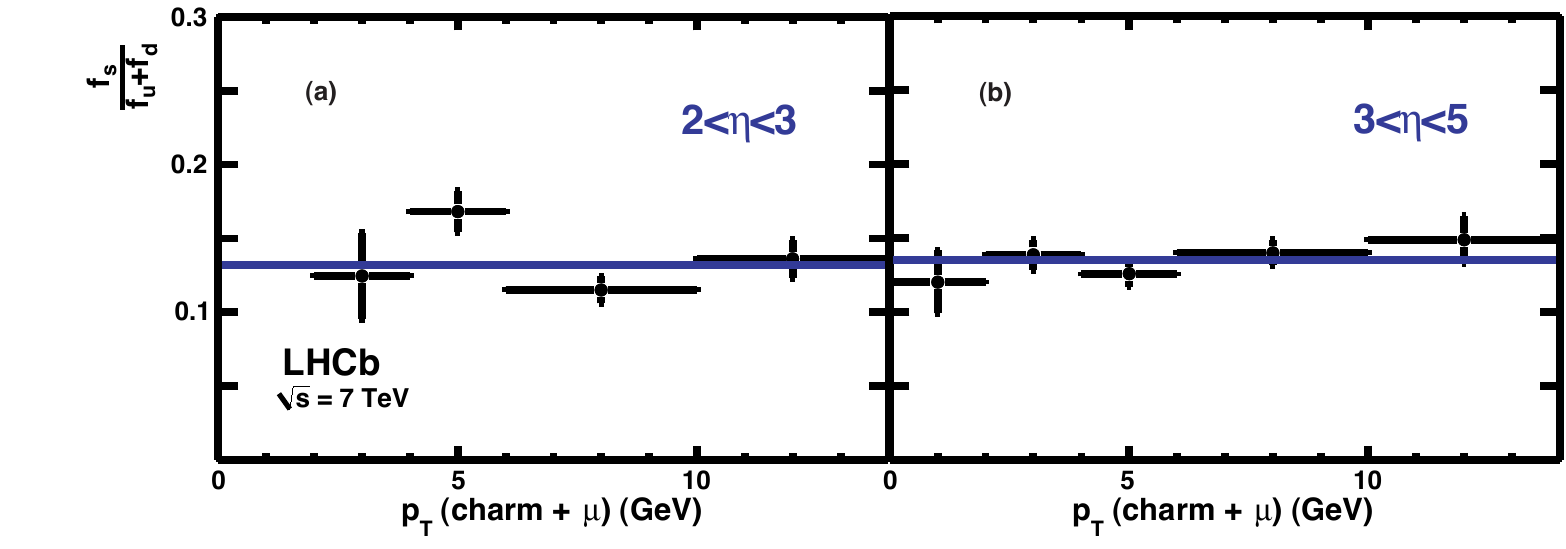}
\end{array}$
\caption{Ratio between $\Bsb$  and light $B$ meson  production fractions as a function of the transverse momentum of the $\Ds\mu^-$ pair in two bins of $\eta$. The errors shown are statistical only.}
\label{fig:fs}\end{center}
\end{figure}

In order to determine the corrected yields entering the ratio $f_s/(f_u+f_d)$, we determine yields in a matrix of three $\eta$ and five $p_{\rm T}$ bins and divide them by the corresponding efficiencies. We then use Eq.~\ref{eq:frac-final}, with the measured lifetime ratio $(\tau_{B^-}+\tau_{\Bzb})/2\tau_{\Bsb}  = 1.07\pm0.02$~\cite{PDG} to derive the ratio $f_s/(f_u+f_d)$ in  two $\eta$ bins. The measured ratio is  constant over the whole $\eta$-$p_{\rm T}$ domain. Figure~\ref{fig:fs} shows the $f_s/(f_u+f_d)$ fractions in bins of $p_{\rm T}$ in two $\eta$ intervals.

\begin{table}[hbt]
\begin{center}
\caption{Systematic uncertainties on the relative $\Bsb$ production fraction.}\label{tab:syserr}
\begin{tabular}{lc}\hline\hline
Source & Error (\%)\\
 \hline
 Bin-dependent errors & 1.0 \\
 ${\cal B}(D^0\to K^-\pi^+)$ & 1.2\\
${\cal B}(D^+\to K^-\pi^+\pi^+)$ & 1.5\\
${\cal B}(\Ds\to K^-K^+\pi^+)$ & 4.9 \\ 
$\Bsb$ semileptonic decay modelling& 3.0\\
Backgrounds & 2.0\\
Tracking efficiency & 2.0 \\
Lifetime ratio & 1.8\\
PID efficiency & 1.5\\
$\Bsb\to D^0 K^+X\munu$ & $^{+4.1}_{-1.1}$\\
${\cal B}((B^-,\Bzb)\to D_s^+KX\munu)$ & 2.0\\
\hline
Total & $^{+8.6}_{-7.7}$\\
\hline\hline
\end{tabular}
\end{center}
\end{table}
\begin{figure}[hbt]
\centering
\includegraphics[width=6 in]{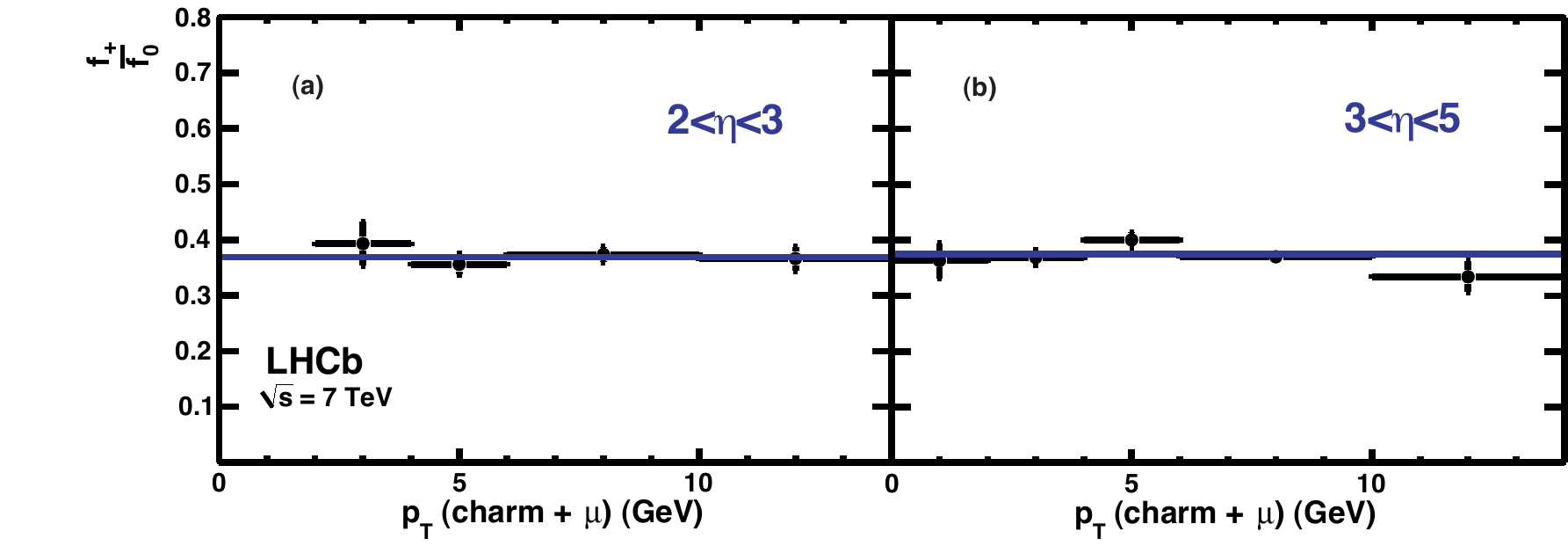}
\caption{$f_+/f_0$ as a function of $p_{\rm T}$ for $\eta$=(2,3) (a) and $\eta$=(3,5) (b). The horizontal line shows the average value.  The error shown combines  statistical and systematic uncertainties accounting for the detection efficiency and the particle identification efficiency.
 } \label{fplus-check}
\end{figure}
By fitting a single constant to all the data, we obtain $f_s/(f_u+f_d)=0.134\pm0.004^{+0.011}_{-0.010}$ in the interval $2<\eta<5$, where the first error is statistical and the second is systematic. The latter  includes several different sources listed in Table~\ref{tab:syserr}.   The dominant  systematic uncertainty  is caused by the experimental uncertainty on ${\cal{B}}(D_s^+\to K^+K^-\pi^+)$
of 4.9\%. 
Adding in the contributions of the $D^0$ and $D^+$ branching fractions we have a systematic error of 5.5\%  due to the charmed hadron branching fractions. The $\Bsb$ semileptonic modelling error is derived by changing the ratio between different hadron species in the final state obtained by removing the SU(3) symmetry constrain, and changing the shapes of the less well known $D^{**}$ states. The tracking efficiency errors mostly cancel in the ratio since we are dealing only with combinations of three or four tracks.  The lifetime ratio error reflects the present experimental accuracy~\cite{PDG}. We correct both for the  bin-dependent PID efficiency obtained with the procedure detailed before, accounting for the statistical error of the calibration sample, and the overall PID efficiency uncertainty, due to the sensitivity to the event multiplicity. The latter is derived by taking the kaon identification efficiency obtained with the method described before, without correcting for the different track multiplicities in the calibration and signal samples. This is compared with the results of the same procedure performed correcting for the ratio of multiplicities in the two samples. The error due to $\Bsb\to D^0 K^+X\munu$ is obtained by changing the  RS/WS background ratio predicted by the simulation within errors, and evaluating the corresponding change in $f_s/(f_u+f_d)$. Finally, the error due to $(B^-,\Bzb)\to D_s^+KX\munu$ reflects the uncertainty in the measured branching fraction.

Isospin symmetry implies the equality of $f_d$ and $f_u$, which allows us to compare $f_+/f_0\equiv n_{\rm corr}(D^+\mu)/n_{\rm corr}(D^0\mu)$ with its expected value. It is not possible to decouple the two ratios for an independent determination of $f_u/f_d$. Using all the known semileptonic branching fractions~\cite{PDG}, we estimate the expected relative fraction of the $D^+$ and $D^0$ modes from $B^{+/0}$  decays to be   $f_+/f_0= 0.375\pm 0.023,$ where the error includes a  6\% theoretical uncertainty associated to the extrapolation of present experimental data needed to account for the inclusive $b\to c \munu$ semileptonic rate.  Our corrected yields correspond to $f_+/f_0=0.373\pm 0.006$ (stat) $\pm$ 0.007 (eff) $\pm$  0.014, for a total uncertainty of 4.5\%.  The last error accounts for uncertainties in $B$  background modelling,   in the  $D^0 K^+\munu$ yield,  the $D^0 p\munu$ yield, the $D^0$ and $D^+$ branching fractions, and tracking efficiency.  The other systematic errors mostly cancel in the ratio. Our measurement of $f_+/f_0$ is not seen to be dependent upon $p_{\rm T}$ or $\eta$, as shown in Fig.~\ref{fplus-check}, and is in agreement with expectation.


\begin{figure}[bt]
\begin{center}
\includegraphics[width=6 in]{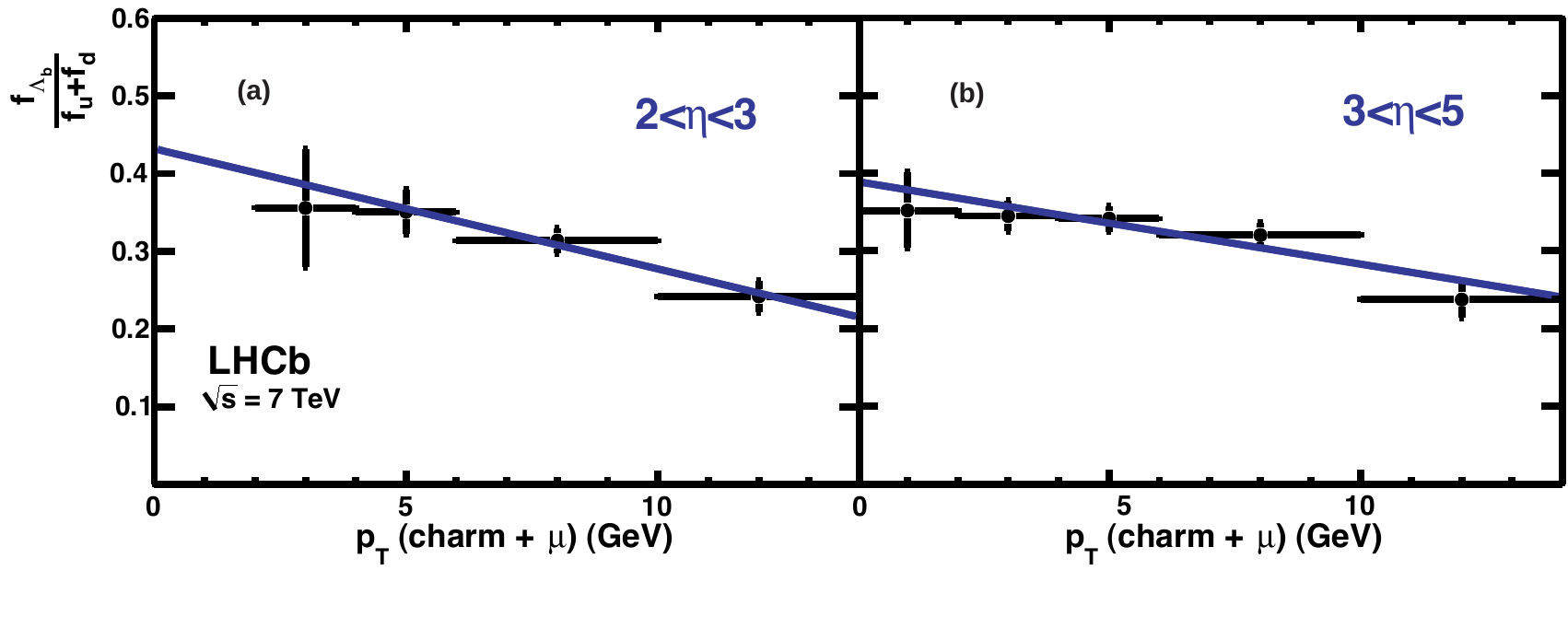}
\caption{Fragmentation ratio $f_{\Lb}/(f_u+f_d)$ dependence upon $p_{\rm T}(\Lc^+\mu^-)$. The errors shown are statistical only.} \label{fig:Lambdab}
\end{center}
\end{figure}

We follow the same procedure to derive the fraction $f_{\Lb}/(f_u+f_d)$, using Eq.~\ref{eq:frac-final-lb} and the ratio $(\tau_{B^-}+\tau_{\Bzb})/(2\tau_{\Lb^0} ) = 1.14\pm 0.03$~\cite{PDG}. In this case, we observe a   $p_{\rm T}$  dependence in the two $\eta$ intervals. Figure~\ref{fig:Lambdab} shows the data fitted to a straight line\begin{equation}
\frac{f_{\Lb}}{f_u+f_d} =a[1+b\times p_{\rm T}({\rm GeV})].
\end{equation}

Table \ref{lin-fit} summarizes the fit results. A corresponding fit to a constant shows that a $p_{\rm T}$ independent  $f_{\Lb}/(f_u+f_d)$ is excluded at the level of four standard deviations. The systematic errors reported in Table~\ref{lin-fit} include only the bin-dependent terms discussed above. 

Table~\ref{tab:syserrLB} summarizes all the sources of absolute scale systematic uncertainties, that include several components. Their definitions mirror closely the corresponding uncertainties for the $f_s/(f_u+f_d)$ determination, and are assessed with the same procedures. The term $\Lb\to D^0 pX\munu$ accounts for the uncertainty in the raw $D^0 pX\munu$ yield, and is evaluated by changing the RS/WS background ratio (1.4$\pm$0.2) within the quoted  uncertainty. In addition, an uncertainty of 2\% is associated with the derivation of the semileptonic branching fraction ratios from the corresponding lifetimes, labelled $\Gamma _{\rm sl}$ in Table~\ref{tab:syserrLB}. The uncertainty is derived assigning conservative errors to the parameters affecting the chromomagnetic operator that influences the $B$ meson total decay widths, but not the $\Lb^0$.  By far the largest term is  the poorly known ${\cal B}(\Lc^+\to pK^- \pi ^+$); thus it is quoted separately.

 \begin{table}[hbt]
\begin{center}	
\caption{Coefficients of the linear fit describing the $p_{\rm T}(\Lc^+\mu^-)$ dependence of $f_{\Lb}/(f_u+f_d)$. The systematic uncertainties included are only those associated with the bin-dependent MC and particle identification errors.}\label{lin-fit}
\begin{tabular}{ccc}\hline\hline
$\eta$ range & $a$ & $b$  \\\hline
2-3 & 0.434$\pm$0.040$\pm$0.025 & -0.036$\pm$0.008$\pm$0.004\\
3-5 & 0.397$\pm$0.020$\pm$0.009 & -0.028$\pm$0.006$\pm$0.003 \\\hline
\hline
2-5 & 0.404$\pm$0.017$\pm$0.009 & -0.031$\pm$0.004$\pm$0.003 \\\hline
\end{tabular}
\end{center}\end{table}

\begin{table}[hbt]
\begin{center}
\caption{Systematic uncertainties on the absolute scale of $f_{\Lambda_b}/(f_u+f_d)$.}\label{tab:syserrLB}
\begin{tabular}{lc}\hline\hline
Source & Error (\%)\\
 \hline
Bin dependent errors & 2.2\\
${\cal B}(\Lb^0\to D^0 pX\munu)$ & 2.0 \\
Monte Carlo modelling & 1.0\\
Backgrounds & 3.0\\
Tracking efficiency & 2.0 \\
$\Gamma _{\rm sl}$ & 2.0 \\
Lifetime ratio & 2.6 \\
PID efficiency & 2.5 \\
\hline
Subtotal & 6.3 \\
\hline
${\cal B}(\Lc^+\to pK^-\pi^+)$ & 26.0\\
Total & $26.8$\\
\hline\hline
\end{tabular}
\end{center}
\end{table}

In view of the observed dependence upon $p_{\rm T}$, we  present our results as 
\begin{equation}
\left[\frac{f_{\Lb}}{f_u+f_d}\right](p_{\rm T})=(0.404\pm0.017\pm0.027\pm 0.105)\times [1- (0.031\pm 0.004\pm0.003) \times p_{\rm T} {\rm (GeV)}], 
\end{equation}
 where the scale factor uncertainties are statistical, systematic, and the error on ${\cal B}(\Lc\to pK^-\pi ^+)$ respectively.  The correlation coefficient between the scale factor and the slope parameter in the fit with the full error matrix is
$-0.63$.
Previous measurements of this fraction have been made at LEP and the Tevatron~\cite{cdf:fractions}. LEP obtains 0.110$\pm$0.019~\cite{hfag}. This fraction has been calculated by combining  direct rate measurements with  time-integrated mixing probability averaged over an unbiased sample of semi-leptonic $b$ hadron decays.
CDF measures $f_{\Lb}/(f_u+f_d)=0.281\pm 0.012 ^{+0.011+0.128}_{-0.056-0.086}$, where the last error reflects the uncertainty in ${\cal B}(\Lc^+\to pK^-\pi^+)$.\
It has been suggested~\cite{cdf:fractions} 
 that the difference between the Tevatron and LEP results is explained by the different kinematics of the two experiments. The average $p_{\rm T}$ of the $\Lc^+\mu^-$ system is 10 GeV  for CDF,  while  the $b$-jets, at LEP, have $p\approx 40$ GeV. LHCb probes an even lower $b$ $p_{\rm T}$ range, while retaining some sensitivity in the CDF kinematic region. These data are consistent with CDF in the kinematic region covered by both experiments, and indicate  that the baryon fraction is higher in the lower $p_{\rm T}$ region.

\section{\boldmath Combined result for  the production fraction  $f_s/f_d$ from LHCb }

From the study of $b$ hadron semileptonic decays reported above, and assuming isospin symmetry, namely $f_u=f_d$, we obtain
\begin{equation}
\left(\frac{f_s}{f_d}\right)_{\rm sl}=0.268 \pm 0.008 ({\rm stat}) ^{+0.022}_{-0.020} ({\rm syst}),\nonumber
\end{equation}
where the first error is statistical and the second is systematic. 

Measurements of this quantity have also been made by LHCb by using hadronic $B$ meson decays~\cite{fs:had}. The ratio determined using the relative abundances of $\Bsb \to \Dsp\pim$ to $\Bzb \to \Dp\Km$  is
\begin{equation}
\left(\frac{f_s}{f_d}\right)_{\rm h1}= 0.250 \pm 0.024 (\rm stat) \pm 0.017({\rm syst})\pm 0.017({\rm theor}),\nonumber
\end{equation}
while that from the relative abundances of $\Bsb \to \Dsp\pim$ to $\Bzb\to \Dp\pim$~\cite{fs:had} is
\begin{equation}
\left(\frac{f_s}{f_d}\right)_{\rm h2}= 0.256 \pm 0.014 ({\rm stat})\pm 0.019({\rm syst}) \pm 0.026({\rm theor}).\nonumber
\end{equation}
The first uncertainty is statistical, the second systematic and the third theoretical. The theoretical uncertainties  in both cases include  non-factorizable SU(3)-breaking effects and form factor ratio uncertainties. The second ratio is affected by an additional source, accounting for the $W$-exchange diagram in the $\Bzb \to \Dp \pim$ decay.

In order to average these results, we consider the correlations between different sources of systematic uncertainties, as shown in Table~\ref{tab:errors}.  We then utilise a generator of pseudo-experiments, where each independent source of uncertainty is generated as a random variable with Gaussian distribution, except for the component   $\Bsb\to \Dz\Kp\mu^-\neumb X$, which is modeled with a bifurcated Gaussian with standard deviations  equal to the positive and negative errors shown in Table~\ref{tab:errors}. This approach to the averaging procedure is motivated by the goal of proper treatment of  asymmetric errors~\cite{barlow}. We assume that the theoretical errors have a Gaussian distribution. 

\begin{table}[hbt]
\caption{Summary of the systematic and theoretical uncertainties in the three LHCb measurements of $f_s/f_d$.}\label{tab:errors}
\begin{center}
\begin{tabular}{lcccl}
\hline\hline
Source & \multicolumn{3}{c} {Error (\%)} & ~\\
~ & $(f_s/f_d)_{\rm sl}$ & $(f_s/f_d)_{\rm h1}$ & $(f_s/f_d)_{\rm h2}$  & ~\\\hline
Bin dependent error & 1.0 & - & - & Uncorrelated \\
Semileptonic decay modelling & 3.0 & - & - & Uncorrelated\\
Backgrounds & 2.0 & - & - & Uncorrelated\\
Fit model & - &2.8 & 2.8& Uncorrelated \\
Trigger simulation & - & 2.0 & 2.0 & Uncorrelated\\
Tracking efficiency & 2.0 & - & -& Uncorrelated \\
${\cal B}(\Bsb \to \Dz\Kp X \munu)$ & $^{+4.1}_{-1.1}$ & - & -  & Uncorrelated\\
 ${\cal B}(\Bzb/B^-\to \Ds KX\munu)$ & 2.0  & - & - & Uncorrelated \\
Particle identification calibration & 1.5 & 1.0 & 2.5 & Correlated \\
$\B$ lifetimes  & 1.5 & 1.5 & 1.5  & Correlated \\
${\cal B}(\Dsp\to \Kp\Km\pip)$ &4.9 &4.9 & 4.9 & Correlated\\
${\cal B}(\Dp\to \Km\pip\pim)$ & 1.5 & 1.5 &1.5 & Correlated\\
SU(3) and form factors & - & 6.1 & 6.1 & Correlated  \\
$W$-exchange & - & - &  7.8 & Uncorrelated  \\
\hline\hline
\end{tabular}
\end{center}
\end{table}

We define the average fraction  as
\begin{equation}
  f_s/f_d = \alpha_1( f_s/f_d)_{\rm sl} + \alpha _2 (f_s/f_d)_{\rm h1} +\alpha _3( f_s/f_d)_{\rm h2},
 \end{equation}
 where
 \begin{equation}
 \alpha _1+ \alpha _2+\alpha _3 =1.
 \end{equation}
The RMS value of $f_s/f_d $ is then evaluated as a function of $\alpha _1$ and $\alpha_2$. 

 We derive the most probable value $f_s/f_d$  by determining the coefficients $\alpha _i$ at which the RMS is minimum, and the total errors by  computing the boundaries defining the 68\% CL, scanning from top to bottom along the axes $\alpha_1$ and $\alpha _2$ in the range comprised between 0 and 1.  The optimal weights determined with this procedure are $\alpha_1=0.73$, and $\alpha_2=0.14$, corresponding to the most probable value
 \begin{equation}
  f_s/f_d =0.267 ^{+0.021}_{-0.020}.\nonumber
 \end{equation}
The most probable value differs slightly from a simple weighted average of the three measurements because of the  asymmetry of the error distribution in the semileptonic determination. By switching off different components we can assess the contribution of each source of uncertainty. Table~\ref{tab:errorcomp} summarizes the results.
 
 \begin{table}
 \caption{Uncertainties in the combined value of $f_s/f_d $.}\label{tab:errorcomp}
 \begin{center}
 \begin{tabular}{lc}
 \hline\hline
 Source & Error (\%)\\
 \hline
 Statistical &  2.8 \\
 Experimental systematic (symmetric) & 3.3 \\
${\cal B}(\Bsb \to \Dz\Kp X \munu)$  & $^{+3.0}_{-0.8}$\\
 \hline
 ${\cal B}(\Dp\to \Km\pip\pim)$ & 2.2 \\
 ${\cal B}(\Dsp\to \Kp\Km\pip)$  & 4.9 \\
 $B$ lifetimes & 1.5 \\
 ${\cal B}(\Bzb/B^-\to \Ds KX\munu)$ & 1.5 \\
 \hline
Theory  & 1.9\\ 
 \hline
 \hline
 
 \end{tabular}
 \end{center}
 \end{table}

\section{Conclusions}
We measure the ratio of the $\Bsb$ production fraction to the sum of those for $B^-$ and $\Bzb$  mesons $f_s/(f_u+f_d)$ = 0.134$\pm$0.004$^{+0.011}_{-0.010}$, and find it consistent with being independent of $\eta$ and $p_{\rm T}$.  
Our results are more precise than, and in agreement with, previous measurements in different kinematic regions.
We combine the LHCb measurements of the ratio of $\Bsb$ to $\Bzb$ production fractions obtained using $b$ hadron semileptonic decays, and two different ratios of  branching fraction of exclusive hadronic decays to derive $ f_s/f_d = 0.267 ^{+0.021}_{-0.020}$. 
The ratio of the $\Lambda _b^0$ baryon production fraction to the sum of those for  $B^-$ and $\Bzb$  mesons varies with the $p_{\rm T}$ of the charmed hadron muon pair. Assuming a linear dependence up to $p_{\rm T}=14$ GeV, we obtain
\begin{equation}
\frac{f_{\Lb}}{f_u+f_d}=(0.404\pm0.017\pm0.027\pm 0.105)\times [1- (0.031\pm 0.004\pm0.003) \times p_{\rm T} {\rm (GeV)}], 
\end{equation}
where the errors on the absolute scale are statistical, systematic and error on ${\cal B}(\Lc^+\to pK^-\pi^+)$ respectively. No $\eta$ dependence is found. 


\section*{Acknowledgements}

\noindent We express our gratitude to our colleagues in the CERN accelerator
departments for the excellent performance of the LHC. We thank the
technical and administrative staff at CERN and at the LHCb institutes,
and acknowledge support from the National Agencies: CAPES, CNPq,
FAPERJ and FINEP (Brazil); CERN; NSFC (China); CNRS/IN2P3 (France);
BMBF, DFG, HGF and MPG (Germany); SFI (Ireland); INFN (Italy); FOM and
NWO (the Netherlands); SCSR (Poland); ANCS (Romania); MinES of Russia and
Rosatom (Russia); MICINN, XuntaGal and GENCAT (Spain); SNSF and SER
(Switzerland); NAS Ukraine (Ukraine); STFC (United Kingdom); NSF
(USA). We also acknowledge the support received from the ERC under FP7
and the Region Auvergne.

\end{document}